\documentclass[aps,pre,twocolumn,superscriptaddress,longbibliography]{revtex4-2}
\usepackage{amsmath}
\usepackage{amsfonts}
\usepackage{amssymb}
\usepackage{hyperref}
\usepackage{lmodern,dsfont}
\usepackage{graphicx}
\usepackage[usenames,dvipsnames]{xcolor}
\usepackage{bm}
\usepackage[english=american]{csquotes}
\usepackage{xr}

\newcommand{\Av}[1]{\left\langle #1 \right\rangle}

\newcommand{\n}{\nonumber}
\newcommand{\nn}{\nonumber \\}

\newcommand{\grad}{\bm{\nabla}}

\renewcommand{\eqref}[1]{Eq.~(\ref{#1})}

\begin{document}
\author{Andreas Dechant}
\author{Jann van der Meer}
\affiliation{Department of Physics No. 1, Graduate School of Science, Kyoto University, Kyoto 606-8502, Japan}

\title{
Renormalized entropy production for optimal transport in jump processes: \\ Make conservative forces optimal again 
}
\date{\today}

\begin{abstract}
For continuous-space diffusion processes, there is a strong connection between conservative forces and entropy production. For a given time evolution of the system's state, the entropy production is minimized when the system is driven by a unique conservative force. However, this relation does not extend to jump processes on a discrete state space. In this case, the forces that minimize the entropy production are generally nonconservative, this effect is more pronounced far from equilibrium in the presence of high energy barriers.
Here we show that, while conservative forces do not minimize the entropy production for a given time evolution, they are nevertheless uniquely characterized as the minimizer of a quantity we dub the renormalized entropy production.
This work explores the properties this quantity shares with entropy production as well as crucial differences between them. We also discuss the conceptual and physical differences between the corresponding optimization problems in finite time. Our theoretical calculations are illustrated with explicit numerical examples.
\end{abstract}

\maketitle

\section{Introduction}

One of the advantages of stochastic thermodynamics \cite{seki10, peli21, shir23, seif25} as a formalized theory for nonequilibrium thermodynamics on the nanoscale is that physical problems attain a clear mathematical formulation. One question where this bridge to mathematics has proved particularly successful is the finite-time optimization problem of driving a physical system from an initial to a final state while minimizing the required entropy production or work. Although originally formulated for overdamped Langevin dynamics in harmonic potentials \cite{schm07a}, it was later discovered that optimal transport theory \cite{mong81, bena00, vill08} has precisely the tools needed to solve this optimization problem for any potential \cite{aure11,aure12, naka21}. 

Applications of optimal transport theory to continuous systems and, more specifically, to overdamped diffusion remain a subject of significant interest, particularly in the context of finite-time corrections to the Landauer principle \cite{beru12, zulk14, dech19, proe20a, blab23, oika25}. Nevertheless, much of the intuition that applies to continuous systems cannot be immediately transferred to the corresponding discrete case of a Markov jump process. A prominent example is the thermodynamic uncertainty relation (TUR) \cite{bara15, ging16, horo20}, which can be saturated for arbitrary driving strength for an overdamped Langevin particle in the absence of a potential. In the corresponding discrete case, a homogeneous unicycle, however, the TUR cannot be saturated beyond linear response near equilibrium. The natural tighter and saturated bound in this case is formulated in terms of a quantity dubbed pseudo-entropy production \cite{shir21}, which reduces to the entropy production near equilibrium and in the continuum limit.

Another at first glance unrelated difference between overdamped diffusion governed by a Langevin equation and discrete jump processes involves the presence of a natural timescale. Whereas in the former case an inherent timescale proportional to the inverse diffusion constant can be specified without referring to the details of the driving forces themselves, characteristic timescales in the latter case always refer to the transition rates between the states, thus there is no obvious natural way to separate inherent system specifics like its timescales from, e.g., controllable external driving forces. 

From the perspective of optimal transport, this implies that while the optimization problem in the continuous case is well-defined and convex, it allows for trivial solutions with diverging transition rates at vanishing dissipation \cite{mura13, dech22c}. The absence of an obvious timescale to fix to obtain a proper optimization problem has led to several different suggestions, including fixing some of the transition rates \cite{ilke22}, the symmetric parts of the transition rates \cite{mura13, reml21}, the edgewise Onsager coefficients \cite{yosh23} or the activity, i.e., the rate of total transitions in the system \cite{dech22c}. For processes in finite time, there is an additional subtle distinction of the constraints based on whether they are imposed at each point in time or on the time-integrated quantity, and on whether the average over all edges or their value at each individual edge is fixed. In both cases, the latter choice results in a less restrictive condition and has been investigated for the sum over the Onsager coefficients \cite{vu23a}, the time-integrated total activity \cite{dech22c} and generalized averages \cite{naga25}. 

Solutions to the corresponding optimization problems that minimize entropy production can differ vastly in their physical properties. For example, the optimal solution can be realized with conservative forces if a time-integrated average over all edges of some quantity like activity \cite{dech22c}, Onsager coefficents \cite{vu23a} or another generalized mean \cite{naga25} is considered fixed. In contrast, if one constrains a quantity at each point in time and at each edge, the solution to the corresponding optimization problem is conservative if the quantity is chosen to be the edgewise Onsager coefficients \cite{yosh23}, but typically not for other types of constraints \cite{shortpaper}. Fixing the symmetric part of the transition rates is a thermodynamically well-motivated choice of constraint \cite{mura13, reml21, seif25, shortpaper}, which falls into this latter category, i.e., achieving optimality requires nonconservative driving. Near equilibrium and in the limiting case of a continuous system, many of the proposed constraints coincide and the feature that nonconservative driving is optimal also disappears \cite{reml21, shortpaper}. Thus, the relation between nonconservative driving and optimality in far-from-equilibrium situations beckons further investigation.

In this work, we approach this elusive question from a different perspective. Instead of trying to find constraints that result in an optimization problem realized by conservative forces, we examine a parametrization of transition rates in which the dissipation-minimizing solution features nonconservative forces, but now ask whether under a different measure of irreversibility the optimal solution is minimized by conservative forces. The introductory Section \ref{sec-minent} illustrates the setup of Markov jump processes and contrasts this situation with the simpler, more familiar continuous case. Our main result, presented in Section \ref{sec-renorm}, is the identification and explicit construction of such a quantity, which we dub the renormalized entropy production. Its relation to entropy production and the corresponding optimization problems in finite time are discussed and numerically illustrated in Section \ref{sec-finite-time}. In Section \ref{sec-renorm-path}, we construct a fluctuating, trajectory-dependent counterpart of the renormalized entropy production, which offers another perspective for comparisons between this quantity and entropy production. The final concluding Section \ref{sec-discussion} sketches possible perspectives for future research.

\section{Minimum entropy production} \label{sec-minent}

\subsection{Diffusion processes} \label{sec-minent-diff}

In a diffusion process, the state of the system is given by a $d$-dimensional real vector $\bm{x}(t) \in \mathbb{R}^d$.
The probability density $p_t(\bm{x})$ of observing a state $\bm{x}$ at time $t$ follows the Fokker-Planck equation \cite{Risken1986}
\begin{subequations}
\begin{align}
\partial_t p_t(\bm{x}) &= - \grad \cdot \big( \bm{\nu}_t(\bm{x}) p_t(\bm{x}) \big) \\
\bm{\nu}_t(\bm{x}) &= \mu \bm{F}_t(\bm{x}) - (\mu/\beta) \grad \ln p_t(\bm{x}) ,
\end{align} \label{fpe}%
\end{subequations}
where $\bm{\nu}_t(\bm{x})$ is the local mean velocity, which measures the local average motion of the system, and $\mu$ is the mobility.
The local mean velocity is determined by a force term $\bm{F}_t(\bm{x})$, which includes interactions between the system's degrees of freedom and external forces, and a diffusive term $- (\mu/\beta) \grad \ln p_t(\bm{x})$, which is a consequence of the fluctuations in the system due to its contact with a heat bath at inverse temperature $\beta$ (we set the Boltzmann constant to unity here and in the following).
Note that it is customary to refer to both $\bm{x}(t)$ and $p_t(\bm{x})$ as the state of the system, the former representing the configuration of a single realization of the process and the latter representing the ensemble of configurations.

We say that the force $\bm{F}_t(\bm{x})$ at time $t$ is conservative if we can define a potential function $U_t(\bm{x})$ of the configuration $\bm{x}$ such that the forces acting in a given configuration are given by the negative gradient of this potential, $\bm{F}_t(\bm{x}) = - \grad U_t(\bm{x})$. 
If the force is conservative and time-independent, $\bm{F}(\bm{x}) = - \grad U(\bm{x})$, then the system will relax to the Boltzmann-Gibbs equilibrium distribution in the long-time limit, assuming that the potential is sufficiently confining,
\begin{align}
p^\text{eq}(\bm{x}) = Z^{-1} \exp\bigg(- \beta U(\bm{x}) \bigg) \label{bg-diffusion},
\end{align}
with $Z = \int d\bm{y} \ e^{-\beta U(\bm{y})}$.
In the equilibrium state, the local mean velocity vanishes $\bm{\nu}^\text{eq}(\bm{x}) = 0$.

In general, however, the system described by \eqref{fpe} is out of equilibrium, which can be characterized by the rate of entropy production
\begin{align}
\sigma_t = \frac{\beta}{\mu} \int d\bm{x} \ \big\Vert \bm{\nu}_t(\bm{x}) \big\Vert^2 p_t(\bm{x}) \label{entropy-rate-diffusion} .
\end{align}
From this expression for the entropy production rate, it is clear that it is positive whenever there is a nonvanishing local mean velocity, and thus whenever the system is out of equilibrium.
This can be caused by preparing a configuration $p(\bm{x})$ that is not the equilibrium one, by changing the potential $U_t(\bm{x})$ as a function of time (or by adjusting external control parameters that affect the potential), or we can apply a genuinely nonconservative force $\bm{F}^\text{nc}(\bm{x})$ to the system. 
The difference is that, while in the former cases, if we imagine keeping the control parameters fixed, the system will eventually relax back into a thermal equilibrium state. 
In the latter case, the system will remain out of equilibrium even in its long-time steady state, which becomes a so-called nonequilibrium steady state. 
In this case, the system is driven by a nonconservative force, which leads to a nonvanishing steady state local mean velocity,
\begin{subequations}
\begin{align}
0 &= - \grad \cdot \big( \bm{\nu}^\text{st}(\bm{x}) p^\text{st}(\bm{x}) \big) \\
\bm{\nu}^\text{st}(\bm{x}) &= \mu \bm{F}(\bm{x}) - (\mu/\beta) \grad \ln p^\text{st}(\bm{x}) .
\end{align} \label{fpe-steady}%
\end{subequations}
In other words, the stationary state of a system is an equilibrium state with vanishing entropy production only if the force is conservative. Thus, the concept of conservative forces remains useful to characterize different ways in which a system may be driven away from equilibrium.

The entropy production quantifies the degree of irreversibility in the system, which can be made explicit by the relation
\begin{align}
\mathcal{S} &= \int_0^\tau dt \ \sigma_t = D_\text{KL} \big(\mathbb{P}(\Gamma) \Vert \mathbb{P}^\dagger(\Gamma^\dagger) \big) \label{entropy-diffusion} \\
&= \int d\Gamma \ \mathbb{P}(\Gamma) \ln \bigg( \frac{\mathbb{P}(\Gamma)}{\mathbb{P}^\dagger(\Gamma^\dagger)} \bigg) \n .
\end{align}
Here $\Gamma = (\bm{x}(t))_{t \in [0,\tau]}$ is the trajectory of the system in the time interval $[0,\tau]$ and $\mathbb{P}(\Gamma)$ is the probability density of observing this trajectory.
The time-reversed trajectory is defined as $\Gamma^\dagger = (\bm{x}(\tau-t))_{t \in [0,\tau]}$ and $\mathbb{P}^\dagger(\Gamma)$ is the probability density of a trajectory in the system driven by the time-reversed force $\bm{F}^\dagger_t(\bm{x}) = \bm{F}_{\tau-t}(\bm{x})$, starting from the final state $p_\tau(\bm{x})$.
\eqref{entropy-diffusion} states that the entropy production of the process in the time interval $[0,\tau]$ is given by the Kullback-Leibler (KL) divergence between the probability of the forward trajectory in the forward dynamics and the time-reversed trajectory in the time-reversed dynamics.
From this expression, we see that entropy production quantifies how much the dynamics change under time reversal, and vanishes only if the system is in equilibrium and thus reversible.

The physical consequence of nonzero entropy production is that energy is dissipated as heat into the environment.
Generally, it is desirable to minimize the entropy production and therefore the losses via dissipation for a given process.
If the forces are conservative and the control parameters are changed sufficiently slowly, the system remains close to the equilibrium state and dissipation can be made arbitrarily small.
By contrast, in the presence of nonconservative forces, slow driving may actually increase the dissipation during the process, as the system keeps dissipating even without any change in the state.
This suggests that, if our goal is to achieve some change in the state with minimal dissipation, driving slowly and avoiding nonconservative forces is advisable.

In practice however, we typically want to realize a change in the system's state within a finite amount of time and thus cannot avoid dissipation entirely.
If we take the time evolution of the system state $p_t(\bm{x})$ as the desired process, this corresponds to minimizing \eqref{entropy-rate-diffusion} under the constraint \eqref{fpe}.
This problem was investigated in \cite{Maes2014,Dechant2022}; the result is that there exists a unique conservative force $\bm{F}_t^*(\bm{x}) = - \grad U^*_t(\bm{x})$ that leads to the same time evolution as \eqref{fpe}, while minimizing the entropy production rate.
This has two consequences: First, because of the uniqueness of this force, if the dynamics \eqref{fpe} is driven by a conservative $\bm{F}_t(\bm{x}) = - \grad U_t(\bm{x})$, then this force is already optimal.
In other words, there is a one-to-one relation between the time evolution of the probability density $p_t(\bm{x})$ and the time-dependent potential $U_t(\bm{x})$.
Second, if the force $\bm{F}_t(\bm{x})$ is nonconservative, then we can find a conservative force $\bm{F}_t^*(\bm{x})$ with the same time evolution and a strictly smaller entropy production rate $\sigma_t^* \leq \sigma_t$.
This means that, even for finite-time changes, it is sufficient to consider conservative forces for the purpose of minimizing dissipation.
Moreover, while the state of the system is no longer given by the Boltzmann-Gibbs distribution, we still have a one-to-one correspondence between the conservative force and the state in the sense that there exists a unique conservative force $\bm{F}_t(\bm{x})$ corresponding to a given time-evolution of the state \cite{Dechant2022,Dechant2022a}.
In conclusion, both in the steady state and for time-dependent states, a conservative force minimizes the entropy production and is uniquely associated with the probability $p_t(\bm{x})$.

\subsection{Jump processes} \label{sec-minent-jump}

Not all types of dynamics can be described as an overdamped diffusion, for many systems a description in terms of a discrete set of states is more appropriate.
For simplicity, we assume the number of states is finite and label them as $i(t) \in \lbrace 1, \ldots, N \rbrace$, where $i(t)$ denotes the current state at time $t$.
The coupling to the thermal environment then induces random transitions between these states; in many cases, transitions between different states $j$ and $i$ are characterized by transition rates $W_t(i,j)$, which measure the average number of transitions from state $j$ to state $i$ per unit time. The corresponding dynamics is then called a Markov jump process.

The occupation probability $p_t(i)$ of state $i$ at time $t$ evolves according to the master equation
\begin{align}
d_t p_t(i) = \sum_{j = 1}^N \big( W_t(i,j) p_t(j) - W_t(j,i) p_t(i) \big) \label{master}.
\end{align}
With given initial occupation probabilities $p_0(i)$.
Note that the diagonal rates $W_t(i,i)$ do not enter the the time-evolution of the occupation probabilities, in the following we choose $W_t(i,i) = 0$ for convenience of notation.
In the following, we use the short-hand notation for sums
\begin{align}
    \sum_j = \sum_{j = 1}^N, \quad \sum_{i,j} = \sum_{i,j = 1}^N .
\end{align}
The first term of \eqref{master} describes transitions from other states into state $i$ and thus increases the probability to be in state $i$, whereas the second term describes transitions from state $i$ to other states, decreasing the probability of state $i$.
In the following, we assume, firstly, that the state space is connected and, secondly, microscopic reversibility $W_t(i,j) \neq 0 \Leftrightarrow W_t(j,i) \neq 0$. These conditions ensure that starting from a single given state, the system will eventually have a nonzero probability to be in any other state.
We also assume that the transition rates can be written in the form \cite{mura13, reml21, seif25}
\begin{align}
W_t(i,j) = \omega_t(i,j) \exp \bigg(-\frac{\beta}{2} A_t(i,j) \bigg) \label{rates}.
\end{align}
Here $\omega_t(i,j) = \omega_t(j,i) \geq 0$ is a symmetric matrix with nonnegative entries; and we set $\omega(i,i) = 0$. If states $i$ and $j$ are connected, then $\omega_t(i,j) \neq 0$, in which case $\omega_t(i,j)^{-1}$ characterizes the time scale of transitions between these states.
By contrast, $A_t(i,j) = - A_t(j,i)$ is a skew-symmetric matrix, which we interpret as the change in energy upon a transition from state $j$ to state $i$ and which is weighted by the inverse temperature $\beta = 1/T$.
The physical interpretation of \eqref{rates} is that transitions which decrease energy are more likely; the parameters $A_t(i,j)$ then are the forces driving the transitions between different states.
We remark that any rates that satisfy the condition $W_t(i,j) = 0 \Leftrightarrow W_t(j,i) = 0$ can be written in the form of \eqref{rates} by choosing
\begin{subequations}
\begin{align}
\omega_t(i,j) &= \sqrt{W_t(i,j) W_t(j,i)} \\
-\beta A_t(i,j) &= \ln \bigg(\frac{W_t(i,j)}{W_t(j,i)} \bigg), \label{local-db}
\end{align}
\end{subequations}
which provides an energetic interpretation of the transition rates in \eqref{master}.
The relation \eqref{local-db} between $A_t(i,j)$ and the logarithm of the transition rates is called the local detailed balance condition; its interpretation is that, whenever a transition from $j$ to $i$ occurs, we associate this with a heat transfer of $-A_t(i,j)$ to the bath and thus an entropy increase of the bath by $-\beta A_t(i,j)$.
It will also be useful to introduce another parametrization of \eqref{master} by defining
\begin{subequations}
\begin{align}
\Omega_t(i,j) &= \omega_t(i,j) \sqrt{p_t(i) p_t(j)} \\
F_t(i,j) &= -\beta A_t(i,j) + \big(\ln p_t(j) - \ln p_t(i) \big) .
\end{align} \label{param}%
\end{subequations}
From the definition, it is clear that $\Omega_t(i,j)$ defines a symmetric nonnegative matrix, while the parameters $F_t(i,j)$, which are called generalized thermodynamic forces, form a skew-symmetric matrix.
From the local detailed balance condition, $F_t(i,j)$ can be interpreted as the change in total (system and bath) entropy upon a transition from $j$ to $i$.
Further, \eqref{master} can be written as
\begin{align}
d_t p_t(i) = 2 \sum_{j} \Omega_t(i,j) \sinh\bigg(\frac{F_t(i,j)}{2} \bigg) \label{master-param} .
\end{align}

Just like for a diffusion process, we define the entropy production for a jump process via \eqref{entropy-diffusion}, where the time-reversal of the protocol is now achieved by choosing the transition rates $W_t^\dagger(i,j) = W_{\tau - t}(i,j)$ in the time-reversed process.
The resulting expression for the entropy production rate is
\begin{align}
\sigma_t = \sum_{i,j} W_t(i,j) p_t(j) \ln \bigg( \frac{W_t(i,j) p_t(j)}{W_t(j,i) p_t(i)} \bigg) \label{entropy-rate-jump} .
\end{align}
The entropy production rate thus quantifies the imbalance between the forward and reverse transition probabilities.
In terms of the parametrization \eqref{param}, this can be written as
\begin{align}
\sigma_t = \sum_{i,j} \Omega_t(i,j) F_t(i,j) \sinh \bigg(\frac{F_t(i,j)}{2} \bigg) \label{entropy-param}.
\end{align}
Since every term in the sum is positive, we see that the entropy production rate only vanishes if for every pair of states $(i,j)$,
\begin{align}
F_t(i,j) = 0 \quad \text{or} \quad \Omega_t(i,j) = 0 .
\end{align}
From \eqref{master-param}, this immediately implies $d_t p_t(i) = 0$ and thus that, naturally, an equilibrium state is always a steady state $p^\text{eq}(i)$.
If $\Omega_t(i,j) = 0$, no transitions between states $j$ and $i$ occur, so in equilibrium, we have $F_t(i,j) = 0$ for every pair of connected states, which from \eqref{param} implies
\begin{align}
\beta A(i,j) = \ln p^\text{eq}(j) - \ln p^\text{eq}(i) .
\end{align}
We can then write the energy change in the transition from $j$ to $i$ as
\begin{align}
A(i,j) = U(i) - U(j) \quad \text{with} \quad U(i) = - \frac{1}{\beta} \ln p^\text{eq}(i) \label{detailed-balance} .
\end{align}
This implies that an equilibrium state is only realized as the steady state in a potential, corresponding to conservative forces in the continuous case.
Thus, both for diffusion and jump processes, we find a one-to-one correspondence between conservative forces and equilibrium states.
As an obvious consequence, among all steady states, the entropy production rate is minimized by the conservative force \eqref{detailed-balance}.
We remark that the condition \eqref{detailed-balance} is independent of the symmetric part $\omega_t(i,j)$ of the transition rates; any choice of $\omega_t(i,j)$ leads to the same equilibrium state.
From \eqref{entropy-param}, we also see that $\sigma_t$ is a positive convex functional of the thermodynamic forces $F_t(i,j)$.
Since the equilibrium state corresponds to $F_t(i,j) = 0$, the magnitude of the thermodynamic forces determines the distance of the system from equilibrium.

For a time-dependent state, we still have a one-to-one correspondence between the occupation probabilities $p_t(i)$ and a time-dependent potential $U_t(i)$.
Specifically, for fixed $\omega_t(i,j)$ and initial state $p_0(i)$, the time evolution of $p_t(i)$ can be realized by a conservative force $A_t(i,j) = U_t(j) - U_t(i)$, where the potential $U_t(i)$ is unique up to a constant shift independent of $i$.
The uniqueness of the potential is explicitly shown in Appendix \ref{app-unique}.
However, we stress that this correspondence only holds when the symmetric part $\omega_t(i,j)$ of the transition rates is held fixed; in general, there are infinitely many combinations of $\omega_t(i,j)$ and $U_t(i)$ that lead to the same time evolution.
We remark that in the diffusion case, the uniqueness of the conservative force also holds only for given parameters $\mu$ and $\beta$, while other combinations of $\mu$, $\beta$ and $U_t(\bm{x})$ may result in the same time evolution.

When minimizing the entropy production \eqref{entropy-rate-jump} for a given time evolution $p_t(i)$ the problem is well-defined only if we introduce additional constraints.
As was discussed in Refs.~\cite{mura13, vu21, dech22c} we may make the entropy production arbitrarily small by choosing increasingly large transition rates with vanishing asymmetry.
Given the relation between conservative forces and the time evolution of $p_t(i)$ discussed above, a natural choice is to fix the symmetric part of the transition rates and minimize the entropy production rate with respect to the forces $A_t(i,j)$ as was done in Refs.~\cite{mura13, reml21}.
However, the surprising result is that the force minimizing \eqref{entropy-rate-jump} is generally not conservative, that is, it cannot be written as the difference of a potential $A_t(i,j) = U_t(j) - U_t(i)$.
Note that since, for fixed occupation probabilities, we have a one-to-one relation between $A_t(i,j)$ and $F_t(i,j)$, see \eqref{param}, the problem of finding the minimum entropy production rate for a given time evolution is equivalent to minimizing \eqref{entropy-param} with respect to $F_t(i,j)$ under the constraint \eqref{master-param}.
As a result, we find the condition
\begin{align}
\frac{F^*_t(i,j)}{2} + \tanh\bigg(\frac{F^*_t(i,j)}{2} \bigg) = \lambda_t(j) - \lambda_t(i) \label{c-optimal-entropy},
\end{align}
where $\lambda_t(i)$ is a Lagrange multiplier.
While the function on the left-hand side is a bijection and the parameters $F^*_t(i,j)$ are uniquely determined by the \enquote{potential landscape} $\lambda_t(i)$, the forces $F^*_t(i,j)$ (and thus $A^*_t(i,j)$) themselves cannot be written as a potential difference \cite{reml21}.
If the system is in a steady state, then the minimum entropy production is zero, corresponding to \eqref{detailed-balance}.
If the system is close to a steady state, then we expect the minimum entropy production to be small and thus, via \eqref{entropy-param}, the optimal values of the parameters $F_t^*(i,j)$ to be likewise small.
In this case, the function of the left-hand side of \eqref{c-optimal-entropy} is almost linear, $x/2 + \tanh(x/2) \simeq x$, and the force is close to a conservative force.

We remark that jump processes are more general than diffusion processes; the latter can generally be obtained as the continuum limit of a suitably defined jump process. In this limit, the force between any two states $F_t^*(i,j)$ becomes small, thus the optimal force becomes conservative holds in the continuum limit \cite{reml21}, which recovers the result for a diffusion process.

\section{Renormalized entropy production and conservative forces} \label{sec-renorm}

Although there is a unique conservative force associated with a given time evolution of a jump process, this force does not minimize the entropy production rate.
Since a conservative force is associated with minimum entropy production in the diffusion case, this poses the question whether it has a similar characterization in terms of a different quantity in the Markov jump case.
In this section, this quantity is introduced as the ``renormalized entropy production''.

\subsection{Definition and construction of the renormalized entropy production rate} \label{app-diff}

We start by introducing the instantaneous probability current $J_t(i,j)$ from state $j$ to state $i$ at time $t$ as
\begin{subequations}
\begin{align}
J_t(i,j) & = W_t(i,j) p_t(j) - W_t(j,i) p_t(i) \\
 & = 2 \, \Omega_t(i,j) \sinh \bigg(\frac{F_t(i,j)}{2} \bigg)
,\end{align} \label{def-current}%
\end{subequations}
which enables us to write the entropy production rate \eqref{entropy-rate-jump} as
\begin{align}
    \sigma_t = \frac{1}{2} \sum_{i, j} J_t(i,j) F_t(i,j)
.\end{align}
Similar to \eqref{def-current}, we introduce the traffic $T_t(i,j)$ between states $i$ and $j$ at time $t$ as
\begin{subequations}
\begin{align}
T_t(i,j) & = W_t(i,j) p_t(j) + W_t(j,i) p_t(i) \\
 & = 2 \, \Omega_t(i,j) \cosh \bigg(\frac{F_t(i,j)}{2} \bigg)
,\end{align} \label{def-traffic}%
\end{subequations}
which is a time-symmetric measure of the rate of jumps along the edge connecting $i$ and $j$ in either direction. From \eqref{def-current}, we see that $T_t(i,j) \geq 2 \Omega_t(i,j)$ with equality only in equilibrium, for this reason $\Omega_t(i,j)$ is also referred to as reduced traffic \cite{mura13}.

Under an imagined small shift of the adjustable parameters $A_t(j,i)$ or, equivalently, $F_t(j,i)$ (cf. \eqref{param}) at a fixed point in time $t$, the differential of entropy production becomes
\begin{align}
    d \sigma _t & = \frac{1}{2} \sum_{i,j} \left(  dJ_t(i,j) F_t(i,j) + J_t(i,j) d F_t(i,j) \right) \nonumber \\
    & = \frac{1}{2} \sum_{i,j} dJ_t(i,j) \left( F_t(i,j)  + 2 \, \frac{J_t(i,j)}{T_t(i,j)} \right)
    \label{eq:rho_differential:dsigma}
.\end{align}
In passing to the second line, we use $d F_t(i,j) = \frac{\partial F_t(i,j)}{\partial J_t(i,j)} d J_t(i,j)$ as well as \eqref{def-current} and \eqref{def-traffic}, which together imply $\frac{\partial J_t(i,j)}{\partial F_t(i,j)} = T_t(i,j)/2$.

Let us consider an infinitesimal change of the driving forces that increases the cycle current $J_{\mathcal{C}}$ along cycle $\mathcal{C}$ but leaves all other currents and therefore also the occupation probabilities invariant. Technically, a cycle $\mathcal{C}$ is defined as a closed, directed loop without self-crossings, i.e., a sequence of directed edges in the network entering each state at most once for which the destination of the last edge coincides with the origin of the first one \cite{schn76a, hill89, jian04}. We specify $d J_t(i,j) = \chi^{\mathcal{C}}_{ij} dJ_{\mathcal{C}}$ with
\begin{align}
    \chi^{\mathcal{C}}_{ij} = \begin{cases}
        1 & \text{ if } j \to i \text{ is in cycle } \mathcal{C} \\
        - 1 & \text{ if } i \to j \text{ is in cycle } \mathcal{C} \\
        0 & \text{ else }
    \end{cases}
,\end{align}
in which case the infinitesimal change in the entropy production rate becomes
\begin{align}
    d \sigma_t & = dJ_{\mathcal{C}} \sum_{ji \in \mathcal{C}} \left( F_t(i,j) + 2 \, \frac{J_t(i,j)}{T_t(i,j)} \right) \nonumber \\
    & = dJ_{\mathcal{C}} \sum_{ji \in \mathcal{C}} \left( F_t(i,j) + 2 \tanh \frac{F_t(i,j)}{2} \right)
    \label{eq:rho_differential:dsigma_c}
.\end{align}
Note that the sum $ji \in \mathcal{C}$ avoids double-counting by including only the direction of an edge that aligns with $\mathcal{C}$.
We can readily see that the condition of a conservative force along the cycle $\mathcal{C}$, which is equivalent to the condition of vanishing affinity $\sum_{ji \in \mathcal{C}} F_t(i,j) = 0$, does not coincide with the condition $d \sigma_t = 0$. 

If, however, we consider a quantity $\tilde{\sigma}$ characterized by the differential
\begin{align}
    d \tilde{\sigma}_t = dJ_{\mathcal{C}} \sum_{ji \in \mathcal{C}} F_t(i,j)
\end{align}
instead, its stationary point is always attained for conservative driving. 

This equation only specifies the differential $d \tilde{\sigma}_t$; to fully characterize $\tilde{\sigma}_t$ we have to specify an integration constant. A natural choice is to demand $\tilde{\sigma}_t = 0$ when $\sigma_t = 0$. This is possible because, as discussed in Section \ref{sec-minent-jump}, any given probability distribution satisfying $p_i > 0$ for all $i$ admits an energy profile for which it is the equilibrium configuration without changing the symmetric part of the rates. This equilibrium configuration is obtained for the set of transition rates satisfying \eqref{detailed-balance}.

We define a function $\tilde{\sigma}$ that satisfies $\tilde{\sigma} = 0$ if all $\phi_{ij} = 0$ and is otherwise specified by its differential
\begin{equation}
    d \tilde{\sigma}_t = \frac{1}{2} \sum_{i,j} F_t(i,j) d J_t(i,j)
    \label{eq:rho_differential:dsigmatilde}
.\end{equation}
The difference $\sigma - \tilde{\sigma}$ is calculated by integrating $d \sigma - d \tilde{\sigma}$. Comparing \eqref{eq:rho_differential:dsigma} to \eqref{eq:rho_differential:dsigmatilde}, we use the shorthand $J_t(F_t'(i,j)) = J_t(i,j, F_t'(i,j))$, to obtain
\begin{align}
    \sigma_t - \tilde{\sigma}_t & = \frac{1}{2} \sum_{i,j} \int_0^{F_t(i,j)} J_t(F_t'(i,j)) d F_t'(i,j) \nonumber \\
    & = \frac{1}{2} \sum_{i,j} \int_0^{F_t(i,j)} 2 \, \Omega_t(i,j) \sinh \bigg(\frac{F'_t(i,j)}{2} \bigg) d F_t'(i,j) \nonumber \\
    & = \sum_{i,j}  \Omega_t(i,j) \left[ 2 \cosh \bigg(\frac{F'_t(i,j)}{2} \bigg) \right]^{F'_t(i,j) = F_t(i,j)}_{F'_t(i,j) = 0} \nonumber \\
    & = \sum_{i,j} T_t(i,j) - 2 \sum_{i,j} \Omega_t(i,j)
    \label{eq:rho_differential:difference}
\end{align}
after using \eqref{def-current} and \eqref{def-traffic} for the second and fourth line, respectively. We are now in a position to define the renormalized entropy production rate $\rho_t$ as
\begin{equation}
    \rho_t := 2 \tilde{\sigma}_t  = 2 \sigma_t - 2 \sum_{i,j} T_t(i,j) + 4 \sum_{i,j} \Omega_t(i,j) \label{renorm-entropy-rate}
.\end{equation}
Note that the quantity defined as $\rho$ in Ref. \cite{shortpaper} is precisely $\tilde{\sigma}_t$, i.e., it differs from $\rho_t$ as identified here by a factor of $2$. The reason for this cosmetic change will become apparent as we explore the properties of $\rho_t$ in the following.

\subsection{Comparison of renormalized entropy production to entropy production}

The previous expression, \eqref{renorm-entropy-rate}, can be combined with \eqref{entropy-param} and \eqref{def-traffic} to recast the renormalized entropy production rate $\rho_t$ as
\begin{align}
\rho_t &=  \sigma_t + \sum_{i,j} \Omega_t(i,j) g \big( F_t(i,j) \big) \quad \text{with} \label{eta} \\
g(x) &= x \sinh\Big(\frac{x}{2}\Big) - 4 \Big( \cosh\Big(\frac{x}{2}\Big) - 1 \Big) , \n
\end{align}
which also motivates the definition of renormalized entropy production as its time-integral,
\begin{align}
\mathcal{R} = \int_0^\tau dt \ \rho_t \label{renorm-entropy} .
\end{align}
The function $g(x)$ is positive and convex, so that $\rho_t \geq 0$ with equality only in equilibrium.
Thus, like the entropy production rate, $\rho_t$ is a convex measure of the distance of the system from equilibrium.
Further, the function $g(x)$ satisfies the inequalities
\begin{align}
0 \leq g(x) \leq x \sinh\Big(\frac{x}{2}\Big).
\end{align}
This translates into the relation between $\rho_t$ and the entropy production rate,
\begin{align}
\sigma_t \leq \rho_t \leq 2 \sigma_t \label{eta-sigma-inequality} .
\end{align}
We see that the renormalized entropy production rate is bounded both from below and above by the entropy production rate.
We remark that the lower bound is tight when $F_t(i,j)$ is small, thus to leading order $\rho_t$ is equal to the entropy production when the system is close to equilibrium.
More precisely, we have 
\begin{align}
\rho_t - \sigma_t \simeq \frac{1}{96}\sum_{i,j} \Omega_t(i,j) F_t(i,j)^4 + O(F_t(i,j)^6)
\end{align}
for small $F_t(i,j)$, while both $\rho_t$ and $\sigma_t$ are of order $F_t(i,j)^2$.
In this sense, $\rho_t$ reduces to $\sigma_t$ near equilibrium, whereas it renormalizes the entropy production far from equilibrium.
In the continuum limit, the thermodynamic forces are likewise small and the renormalized entropy production coincides with the entropy production.

We can obtain a more direct interpretation of $\rho_t$ by expressing its definition, \eqref{renorm-entropy-rate}, as
\begin{align}
\rho_t = \sigma_t  + \Big[ \sigma_t - \big(4 \mathcal{A}_t - 4 \mathcal{A}_t^0 \big) \Big],
\end{align}
where we define the dynamical activity
\begin{align}
\mathcal{A}_t = \frac{1}{2} \sum_{i,j} T_t(i,j),
\end{align}
which measures the overall rate at which transitions occur in the system, and the reference activity
\begin{align}
    \mathcal{A}_t^0 = \sum_{i,j} \Omega_t(i,j),
\end{align}
which corresponds to a system with the same symmetric part of the transition rates but vanishing thermodynamic forces. The difference
$\mathcal{A}_t - \mathcal{A}_t^0 \geq 0$ can be interpreted as an excess activity of the system, measuring how much the dynamical activity is increased by the presence of nonzero thermodynamic forces.
Since the term in square brackets is positive, we find that
\begin{align}
    0 \leq \mathcal{A}_t - \mathcal{A}_t^0 \leq \frac{1}{4} \sigma_t ,
\end{align}
so that the excess activity is bounded from above by the entropy production rate.
While both $\mathcal{A}_t$ and $\mathcal{A}_t^0$ diverge in the continuum limit, the excess activity and hence also $\rho_t$ remains finite.
Moreover, the upper bound becomes an equality in the limit of vanishing thermodynamic forces; near equilibrium or in the continuum limit, the entropy production rate coincides with the excess activity.
These relations motivate the denomination of $\rho_t$ as ``renormalized'' entropy production rate: It reweighs the entropy production rate with an a difference of two activity-dependent terms, which remains finite even in the continuum or large-$N$-limit in which the activity itself become infinite.

Finally, we remark that $\rho_t$ can also be written as
\begin{align}
\rho_t &= \frac{1}{2}\sum_{i,j} J_t(i,j) h\big( F_t(i,j) \big) \label{renorm-current} ,
\end{align}
where we defined $h(x) = 2 (x - 2\tanh(x/4))$.
The analog expression for the entropy production rate is
\begin{align*}
\sigma_t = \frac{1}{2} \sum_{i,j} J_t(i,j) F_t(i,j) .
\end{align*}
Thus, while the entropy production rate measures the currents $J_t(i,j)$ weighted by the thermodynamic forces $F_t(i,j)$, the weighting function $h(x)$ in the renormalized entropy production is non-linear in the thermodynamic forces.
We have $h(x) \simeq x$ for small $x$ and $h(x) \simeq 2 x$ for large $x$, so that the upper bound $\rho_t \simeq 2 \sigma_t$ in \eqref{eta-sigma-inequality} can only be realized if the thermodynamic force across every single transition is large, that is, very far from equilibrium.

\subsection{Minimization of renormalized entropy}
Since, like the entropy production rate, $\rho_t$ measures how far the system is from equilibrium, we may ask for the dynamics that minimizes $\rho_t$ for a given time evolution.
This corresponds to minimizing \eqref{eta} with respect to $F_t(i,j)$ under the constraint \eqref{master-param}.
The result is a condition on the optimal parameters $F_t^*(i,j)$,
\begin{align}
F_t^*(i,j) = \lambda_t(j) - \lambda_t(i)  \label{c-optimal-eta},
\end{align}
where $\lambda_t(i)$ is a Lagrange multiplier, determined by the condition
\begin{align}
d_t p_t(i) = 2 \sum_j \Omega_t(i,j) \sinh \bigg(\frac{\lambda_t(j) - \lambda_t(i)}{2} \bigg) \label{lambda-eq} .
\end{align}
As we discuss below, a solution to \eqref{lambda-eq} always exists and is unique up to a constant shift.
The condition \eqref{c-optimal-eta} implies that the optimal thermodynamic forces $F_t^*(i,j)$ can be written as the difference of a potential, which translates into an expression for the energy differences,
\begin{align}
A^*_t(i,j) &= U^*_t(i) - U^*_t(j) \qquad \text{with} \label{optimal-potential} \\
U^*_t(i) &= \frac{1}{\beta} \big(\lambda_t(i) - \ln p_t(i) \big) \n .
\end{align}
Thus, we find that, in contrast to the entropy production rate, the force minimizing the renormalized entropy production rate $\rho_t$ is conservative.
Since this force is simultaneously the unique conservative force corresponding to a given time evolution and symmetric part of the transition rates, we therefore find that the correspondence between conservative forces and entropy production from the diffusion case is recovered at the level of the renormalized entropy production rate.

Since $\rho_t$ attains its minimal value $\rho_t^*$ for conservative forces, we can use $\rho_t - \rho_t^*$ as a measure of how nonconservative the forces in the system are.
This resembles the decomposition of entropy production into (conservative) excess and (nonconservative) housekeeping contributions \cite{hata01,dech22,yosh23}.
However, for a jump process, conservative forces minimize the entropy production only when, instead of the symmetric part of the rates $\omega_t(i,j)$, the Onsager coefficients are fixed \cite{yosh23},
\begin{align}
    L_t(i,j) = \frac{J_t(i,j)}{F_t(i,j)} = \frac{W_t(i,j) p_t(j) - W_t(j,i) p_t(i)}{\ln \Big( \frac{W_t(i,j) p_t(j)}{W_t(j,i) p_t(i)} \Big)} ,
\end{align}
which have been used to define the so-called state mobility \cite{vu23a}.
Specifically, fixing $L_t(i,j)$ and minimizing the entropy production rate with respect to $F_t(i,j)$ yields conservative forces and we can interpret the difference between the entropy production rate and the minimal value as a nonconservative housekeeping entropy.
On the other hand, if $\omega_t(i,j)$ is kept fixed, the optimal forces are nonconservative and thus the difference between the entropy production and its minimal value is not indicative of the presence of nonconservative forces.
This suggests that, under the constraint of fixed $\omega_t(i,j)$, the decomposition of the renormalized entropy production rate $\rho_t$ is more natural to distinguish between conservative and nonconservative effects.

Finally, we show that the solution to the minimization problem of $\rho_t$ exists and is unique.
To do so, we write
\begin{align}
D_t(i,j) = \sinh\bigg(\frac{F_t(i,j)}{2} \bigg) .
\end{align}
In terms of the parameters $D_t(i,j)$, the optimization problem we have to solve is
\begin{align}
\inf_{D_t} & \bigg( 2 \sum_{i,j} \Omega_t(i,j) D_t(i,j) \text{arsinh}\big( 2 D_t(i,j) \big) \bigg) \\
\text{with} &\qquad d_t p_t(i) = 2 \sum_{j} \Omega_t(i,j) D_t(i,j) \n .
\end{align}
This is the minimization of a convex and coercive functional under linear equality constraints, a standard problem in convex optimization \cite{boyd04}.
Since at least one choice of $D_t(i,j)$ (the one defined by the original rates) satisfies the constraints, the problem is feasible and thus its solution exists and is unique.
Since the relation between $D_t(i,j)$ and $F_t(i,j)$ a bijection, the same is true for the optimal $F^*_t(i,j)$.
Thus, the optimal parameters $F^*_t(i,j)$ minimizing $\rho_t$ exist and are unique, which, via \eqref{c-optimal-eta}, implies the existence of the corresponding potential landscape and its uniqueness up to a constant shift.

\section{Near-optimality of conservative forces for finite-time transport} \label{sec-finite-time}

Up to this point, we focused on the minimization of the instantaneous entropy production and renormalized entropy production rates $\sigma_t$ and $\rho_t$, under the constraint of a given time-evolution of $p_t(i)$.
However, in practice, one may be more interested in the problem of starting from a probability distribution $p_0(i) = p_\text{ini}(i)$ at time $t = 0$ and transporting it into a finial probability distribution $p_\tau(i) = p_\text{fin}(i)$ over a finite time-interval of length $\tau$.
Formally, this problem can be separated into specifying a time-evolution $p_t(i)$ with the correct initial and final state, minimizing the instantaneous rates $\sigma_t$ or $\rho_t$ at any given $t$ and then minimizing the total (renormalized) entropy production with respect to the specified time-evolution. The latter point has already been addressed above and in Ref. \cite{shortpaper}; this section focuses on the minimization problem of finding the optimal time-evolution.

As a consequence, the above findings regarding the conservativeness of the resulting thermodynamic forces remain valid; specifically, minimizing the entropy production under the constraint on the symmetric part of the transition rates will result in non-conservative forces at any instant in time, while minimizing the renormalized entropy production will result in conservative forces. It is now a natural question how the solutions minimizing the (renormalized) entropy production relate to each other, since both $\sigma_t$ and $\rho_t$ measure the degree to which the system is out of equilibrium.

\subsection{Relations for a given time-evolution}

Recalling some of our results from Ref. \cite{shortpaper}, we first discuss the case that the time-evolution is specified. 
In the following, we refer to the dynamics minimizing the entropy production with a superscript ``opt'', while the conservative forces minimizing the renormalized entropy production are denoted by a superscript ``cons''.
Thus $\sigma_t^\text{opt} = \sigma_t^*$ is the minimum entropy production rate, while $\rho_t^\text{opt}$ is the renormalized entropy production rate evaluated with the forces obtained by minimizing $\sigma_t$.
Similarly, we have the entropy production rate $\sigma_t^\text{cons}$ in the conservative dynamics minimizing $\rho_t$ and the minimum value $\rho^\text{cons}_t = \rho_t^*$.
From \eqref{eta-sigma-inequality}, we then have
\begin{subequations}
\begin{align}
\sigma_t^\text{opt} \leq &\rho_t^\text{opt} \leq 2 \sigma_t^\text{opt} \\
\sigma_t^\text{cons} \leq &\rho_t^\text{cons} \leq 2 \sigma_t^\text{cons} .
\end{align}
\end{subequations}
Since $\sigma_t^\text{opt}$ and $\rho_t^\text{cons}$ are the respective minimum values, we further have $\sigma_t^\text{opt} \leq \sigma_t^\text{cons}$ and $\rho_t^\text{cons} \leq \rho_t^\text{opt}$.
This results in the chain of inequalities
\begin{align}
\sigma_t^\text{opt} \leq \sigma_t^\text{cons} \leq \rho_t^\text{cons} \leq \rho_t^\text{opt} \leq 2 \sigma_t^\text{opt} \leq 2 \sigma_t^\text{cons} \label{inequalities-all} .
\end{align}
From this, we can draw two important conclusions.
First, the minimum values of $\sigma_t$ and $\rho_t$ are related by
\begin{align}
\frac{1}{2}\rho_t^\text{cons} \leq \sigma_t^\text{opt} \leq \rho_t^\text{cons},
\end{align}
so that the minimum of the entropy production rate is bounded from below and above by the minimum of the renormalized quantity. 
Thus, minimizing $\rho_t$ also yields a strong constraint on the minimum entropy production rate.
Second, since $\sigma_t^\text{cons}$, which is the entropy production rate under the unique conservative force with the same time evolution, satisfies the inequalities
\begin{align}
\sigma_t^\text{opt} \leq \sigma_t^\text{cons} \leq 2 \sigma_t^\text{opt} \label{entropy-near-optimal} .
\end{align}
This implies that, while it is generally possible to reduce the entropy production rate by introducing nonconservative forces, the reduction is at most by a factor of 2.
Likewise, we have the relation for $\rho_t$,
\begin{align}
\rho_t^\text{cons} \leq \rho_t^\text{opt} \leq 2 \rho_t^\text{cons} \label{renentropy-near-optimal}.
\end{align}
Thus, while the entropy production rate is minimized by nonconservative forces and the renormalized entropy production rate by conservative forces, the forces minimizing either functional result in similar values for the respective other functional.
In particular, conservative forces are nearly optimal regarding the minimization of entropy production: 
We can always choose a conservative protocol, which results in an entropy production that is within a factor 2 of the minimal value and has the additional property that the time-evolutions are identical.

\subsection{Optimizing the time-evolution}

The previous relation compares the optimal nonconservative solution to a related conservative solution which, however, is not optimized. If we wish to compare the optimal nonconservative solution minimizing the entropy production to the optimal conservative one that actually minimizes the renormalized entropy production, the time-evolution of these two solutions will in general differ. In the following, we compare these minimization problems in detail.

\begin{widetext}
Concretely, minimizing the entropy production is achieved by solving the optimization problem
\begin{align}
    \min_{F_t(i,j),p_t(i),\lambda_t(i)} \int_0^\tau dt \ &\sum_{i,j} \omega(i,j) \sqrt{p_t(i) p_t(j)} F_t(i,j) \sinh \bigg( \frac{F_t(i,j)}{2} \bigg) \label{entropy-finite-time} \\ &- \sum_i \lambda_t(i) \Bigg( d_t p_t(i) - 2 \sum_j \omega(i,j) \sqrt{p_t(i) p_t(j)} \sinh \bigg( \frac{F_t(i,j)}{2} \bigg) \Bigg) \n
\end{align}
\end{widetext}
with the constraints $p_0(i) = p_\text{ini}(i)$ and $p_\tau(i) = p_\text{fin}(i)$.
Here, we use that optimizing with respect to the parameters $A_t(i,j)$ is equivalent to optimizing with respect to the thermodynamic forces $F_t(i,j)$.
The Lagrange multipliers $\lambda_t(i)$ ensure that the time-evolution of the probability distribution is generated by the master equation \eqref{master-param}.
For simplicity, we assume that the symmetric part of the transition rates $\omega(i,j)$ does not depend on time.
Taking the variation with respect to $F_t(i,j)$ yields \eqref{c-optimal-entropy}, which determines the thermodynamic forces as a non-linear function of the difference in the \enquote{potential} $\lambda_t(i)$.
We define the function $g(x)$ as the inverse of the function $g^{-1}(x) = x + \tanh(x)$, which yields
\begin{align}
    F_t(i,j) = 2 g(\lambda_t(j) - \lambda_t(i)) .
    \label{entropy-finite-time-force-potential}
\end{align}
Plugging this into \eqref{entropy-finite-time} and minimizing with respect to $p_t(i)$, we then obtain the following coupled differential equations for $p_t(i)$ and $\lambda_t(i)$,
\begin{subequations}
\begin{align}
    d_t p_t(i) &= 2\sum_{j} \omega(i,j) \sqrt{p_t(i) p_t(j)} \sinh \big( g(\lambda_t(j) - \lambda_t(i)) \big) \label{entropy-finite-time-ode-probability} \\
    d_t \lambda_t(i) &= \sum_j \omega(i,j) \sqrt{\frac{p_t(j)}{p_t(i)}} \Big( g(\lambda_t(j) - \lambda_t(i)) \label{entropy-finite-time-ode-potential} \\
    &\qquad- 2(\lambda_t(j) - \lambda_t(i)) \Big) \sinh \big( g(\lambda_t(j) - \lambda_t(i)) \big) \n .
\end{align} \label{entropy-finite-time-ode}%
\end{subequations}
These equations must be solved under the constraint on the initial and final condition on $p_t(i)$, which is equivalent to finding the initial potential $\lambda_0(i)$ that results in the correct final distribution $p_\tau(i) = p_\text{fin}(i)$ starting from $p_0(i) = p_\text{ini}(i)$.
In practice we solve these equations by choosing a candidate initial potential $\tilde{\lambda}_0(i)$ and integrating \eqref{entropy-finite-time-ode}, which results in a final probability distribution $\tilde{p}_\tau(i)$ that generally does not match the desired one.
Next, we choose a slightly displaced initial potential $\tilde{\lambda}_0(i) + \delta \lambda_0(i)$ and again compute the final distribution.
By comparing the two final distributions, we can estimate the Jacobian matrix $J(i,j) = \delta p_\tau(i)/\delta \lambda_0(j)$ and then update the initial potential using the Newton-Raphson method $\tilde{\lambda}_0(i) \rightarrow \tilde{\lambda}_0(i) - \sum_j J^{-1}(j,i) (\tilde{p}_\tau(j) - p_\tau(j))$.
Repeating this procedure until the resulting final distribution is sufficiently close to the desired one, we obtain the correct initial potential.
We note that, multiplying \eqref{entropy-finite-time-ode-potential} by $p_t(i)$ and summing over $i$, we obtain the relation
\begin{align}
    \sigma_t^* = - d_t \sum_i \lambda_t(i) p_t(i)
\end{align}
for the optimal entropy production rate.
Thus, the minimal entropy production rate can be expressed as
\begin{align}
    \label{entropy-finite-time-potential}
    \Sigma^* = \Av{\lambda_0}_0 - \Av{\lambda_\tau}_\tau,
\end{align}
which means that the optimal potential solving \eqref{entropy-finite-time-ode} also acts as a potential for the entropy production such that the latter can be expressed as an average potential difference.

\begin{widetext}
In a similar manner, we can minimize the renormalized entropy production, which results in the optimization problem
\begin{align}
    \min_{F_t(i,j),p_t(i),\lambda_t(i)} \int_0^\tau dt \ &\sum_{i,j} \omega(i,j) \sqrt{p_t(i) p_t(j)} \Bigg( 2 F_t(i,j) \sinh \bigg( \frac{F_t(i,j)}{2} \bigg) - 4 \bigg( \cosh\bigg(\frac{F_t(i,j)}{2} \bigg) - 1 \bigg) \Bigg) \label{renentropy-finite-time} \\ &- \sum_i \lambda_t(i) \Bigg( d_t p_t(i) - 2 \sum_j \omega(i,j) \sqrt{p_t(i) p_t(j)} \sinh \bigg( \frac{F_t(i,j)}{2} \bigg) \Bigg) \n .
\end{align}
\end{widetext}
In contrast to the minimization of the entropy production, the thermodynamic forces are now expressed as a linear function of the potential difference,
\begin{align}
    F_t(i,j) = \lambda_t(j) - \lambda_t(i),
\end{align}
and are thus conservative.
We find the differential equations
\begin{subequations}
\begin{align}
    d_t p_t(i) &= 2\sum_{j} \omega(i,j) \sqrt{p_t(i) p_t(j)} \sinh \bigg( \frac{\lambda_t(j) - \lambda_t(i)}{2} \bigg) \label{renentropy-finite-time-ode-probability} \\
    d_t \lambda_t(i) &= 2 \sum_j \omega(i,j) \sqrt{\frac{p_t(j)}{p_t(i)}} \Bigg( \cosh \bigg( \frac{\lambda_t(j)-\lambda_t(i)}{2} \bigg) -1 \Bigg) \label{renentropy-finite-time-ode-potential} .
\end{align} \label{renentropy-finite-time-ode}%
\end{subequations}
Just as before, we need to find the initial potential $\lambda_0(i)$ that results in the correct final distribution.
For the optimal potential solving \eqref{renentropy-finite-time-ode}, we further have
\begin{align}
    \rho_t^* = - d_t \sum_{i} \lambda_t(i) p_t(i),
\end{align}
which allows us to likewise represent the minimal renormalized entropy production as a difference of the average optimal potential,
\begin{align}
    \mathcal{R}^* = \Av{\lambda_0}_0 - \Av{\lambda_\tau}_\tau .
    \label{renentropy-finite-time-potential}
\end{align}
We remark that while the structure of the optimization problems for the entropy production \eqref{entropy-finite-time-ode} and the renormalized entropy production \eqref{renentropy-finite-time-ode} are similar, from a practical point, the latter is numerically easier to tackle as it does not require evaluating the function $g(x)$ in every step.

Another concern is that the non-conservative forces required for minimizing the entropy production rate may be more difficult to implement in practice.
Thus, we may seek to minimize the entropy production under the additional constraint of conservative forces, $F_t(i,j) = \mu_t(j) - \mu_t(i)$, i.~e.~we search for the set of conservative forces that minimizes the entropy production. 
\begin{widetext}
This leads to the optimization problem
\begin{align}
    \min_{\mu_t(i),p_t(i),\lambda_t(i)} \int_0^\tau dt \ &\sum_{i,j} \omega(i,j) \sqrt{p_t(i) p_t(j)} \big(\mu_t(j) - \mu_t(i) \big) \sinh \bigg( \frac{\mu_t(j) - \mu_t(i)}{2} \bigg) \label{entropy-finite-time-conservative} \\ &- \sum_i \lambda_t(i) \Bigg( d_t p_t(i) - 2 \sum_j \omega(i,j) \sqrt{p_t(i) p_t(j)} \sinh \bigg( \frac{\mu_t(j) - \mu_t(i)}{2} \bigg) \Bigg) \n .
\end{align}
Taking the variation with respect to $\mu_t(i)$ and $p_t(i)$, we obtain the relation
\begin{align}
    &\sum_j \omega(i,j) \sqrt{p_t(i) p_t(j)} \Bigg( 2 \sinh\bigg( \frac{\mu_t(j) - \mu_t(i)}{2} \bigg) + \Big(\big(\mu_t(j) - \mu_t(i) \big) - 2 \big(\lambda_t(j) - \lambda_t(i) \big) \Big) \cosh\bigg( \frac{\mu_t(j) - \mu_t(i)}{2} \bigg) \Bigg) = 0 .
\end{align}
\end{widetext}
While this formally relates the potential $\mu_t(i)$ determining the thermodynamic forces to the Lagrange multipliers $\lambda_t(i)$, it represents a nonlinear system of equations that must be solved numerically in every step and we thus cannot obtain a closed set of equations involving only $\lambda_t(i)$ and $p_t(i)$ like \eqref{entropy-finite-time-ode} or \eqref{renentropy-finite-time-ode}.
As a consequence, determining the conservative forces minimizing the entropy production is much harder than determining the globally optimal non-conservative forces.

Here, however, the renormalized entropy production provides a way out:
Since it is naturally minimized by conservative forces, we do not have to impose any additional constraint and can directly solve \eqref{renentropy-finite-time-ode}.
Moreover, the resulting conservative forces are near-optimal also for the entropy production rate.
More precisely, we have the inequalities
\begin{align}
    \mathcal{R}^* = \mathcal{R}^\text{cons}[p_t^{\text{min}\mathcal{R}}] &\leq \mathcal{R}^\text{cons}[p_t^{\text{min}\Sigma}] \label{entropy-finite-time-conservative-inequalities} \\
    &\leq \mathcal{R}^\text{opt}[p_t^{\text{min}\Sigma}] \leq 2 \Sigma^\text{opt}[p_t^{\text{min}\Sigma}] \leq 2 \Sigma^* . \n
\end{align}
Here, $\mathcal{R}^\text{cons}[p_t^{\text{min}\mathcal{R}}]$ is the renormalized entropy production for conservative forces and for the probability-protocol $p_t^{\text{min}\mathcal{R}}$ that is optimal for the renormalized entropy production, i.e., the global minimum of the renormalized entropy production.
Since $p_t^{\text{min}\mathcal{R}}$ is optimal, any other protocol will result in a larger renormalized entropy production, even if the forces are conservative and thus instantaneously optimal; this is in particular true for the protocol $p_t^{\text{min}\Sigma}$ that is optimal for the entropy production.
Replacing the instantaneously optimal conservative forces by the non-conservative forces minimizing the entropy production rate will result in a yet larger renormalized entropy production $\mathcal{R}^\text{opt}[p_t^{\text{min}\Sigma}]$.
Due to \eqref{inequalities-all}, the renormalized entropy production is upper-bounded by twice the entropy production for the same forces and protocol, which implies the final inequality.
Finally, we also have
\begin{align}
    \Sigma^\text{cons}[p_t^{\text{min}\mathcal{R}}] \leq \mathcal{R}^\text{cons}[p_t^{\text{min}\mathcal{R}}],
\end{align}
since the entropy production rate is smaller than the renormalized entropy production rate for the same forces and protocol.
In summary, we thus have
\begin{align}
    \Sigma^* \leq \Sigma^\text{cons}[p_t^{\text{min}\mathcal{R}}] \leq 2 \Sigma^*,
\end{align}
which extends \eqref{entropy-near-optimal} to finite times.
This implies that the conservative forces and the protocol minimizing the renormalized entropy production (i.~e.~the solution of \eqref{renentropy-finite-time-ode}) result in an entropy production that is at most twice as large as the globally minimal one.
Thus, \eqref{renentropy-finite-time-ode} provides a concrete prescription to obtain a protocol of near-optimal conservative forces $F_t(i,j) = \lambda_t(j) - \lambda_t(i)$ that are guaranteed to be within a factor 2 of the truly optimal, but non-conservative, protocol.

\begin{figure}
    \includegraphics[width=0.48\textwidth]{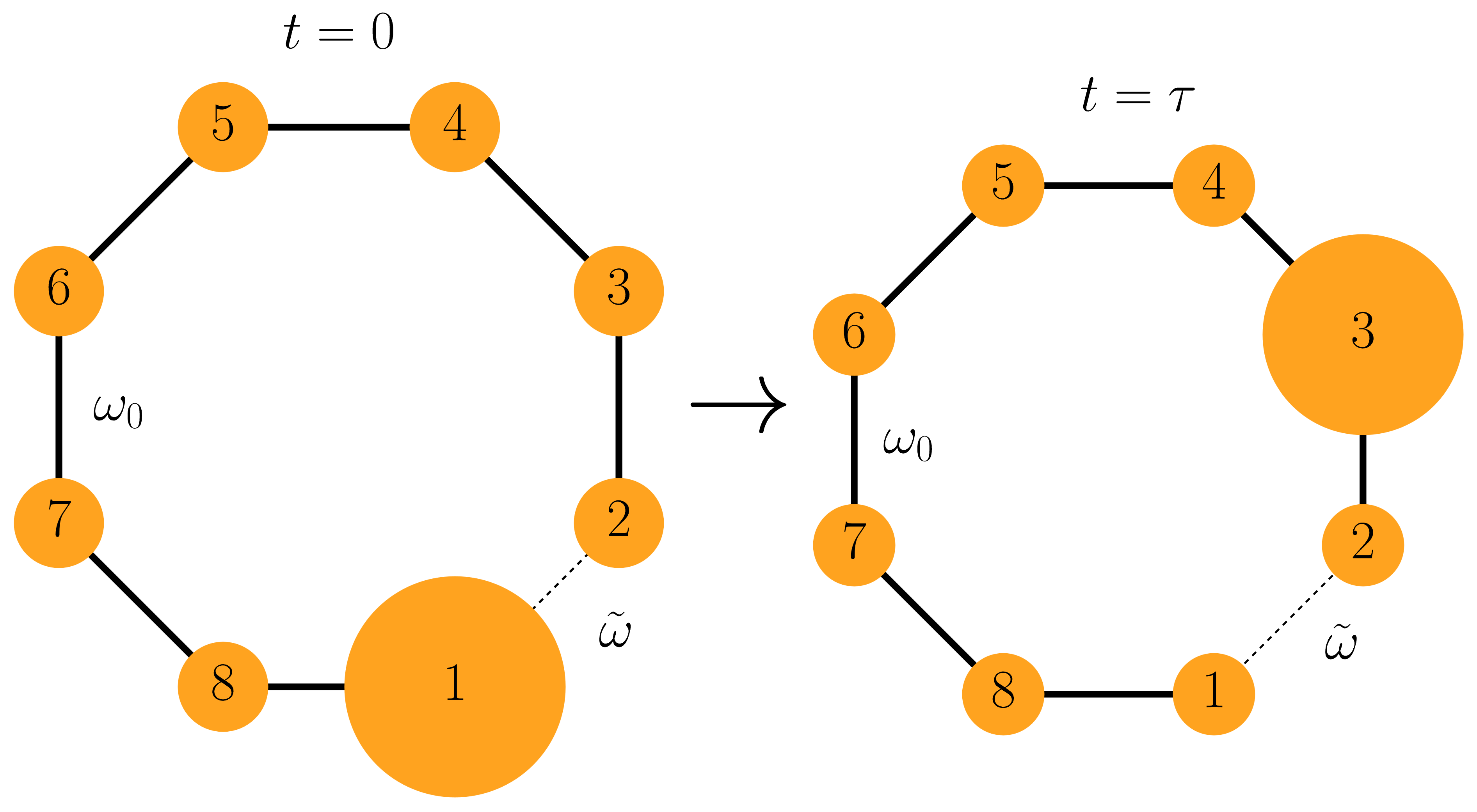}
    \caption{Illustration of the example for a finite-time transport process. The system consists of $N=8$ sites in a ring configuration, with transitions between neighboring sites. The symmetric part of the transition rates is $\omega(i,i+1) = \omega_0$, except for the transition between sites $1$ and $2$, which occurs at the lower rate $\omega(1,2) = \tilde{\omega}$. The size of the orange circles indicates the probability in the initial and final state: Initially, at $t = 0$, the most likely site is site $1$, with all other sites being equiprobable, while finally at $t = \tau$, we want the system to be at site $3$ with the largest probability. This requires us to transport probability from site $1$ to site $3$, which can be done either along the long path through the \enquote{bulk} of sites $4-8$ with large transition rates, or through the weak link connecting site $2$ with its much smaller transition rate.}
    \label{fig-setup}
\end{figure}

\begin{figure}
    \includegraphics[width=0.48\textwidth]{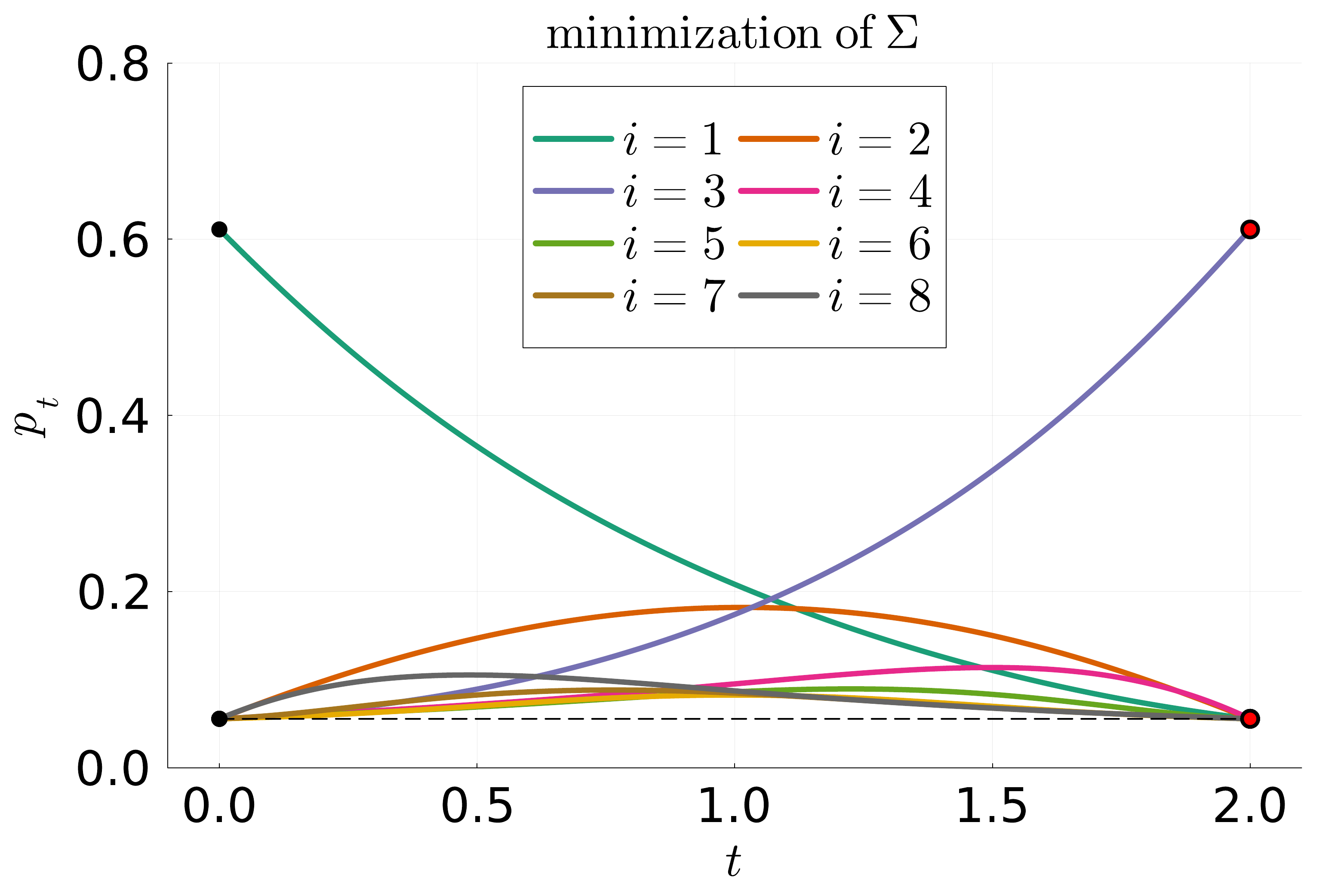}\\
    \includegraphics[width=0.48\textwidth]{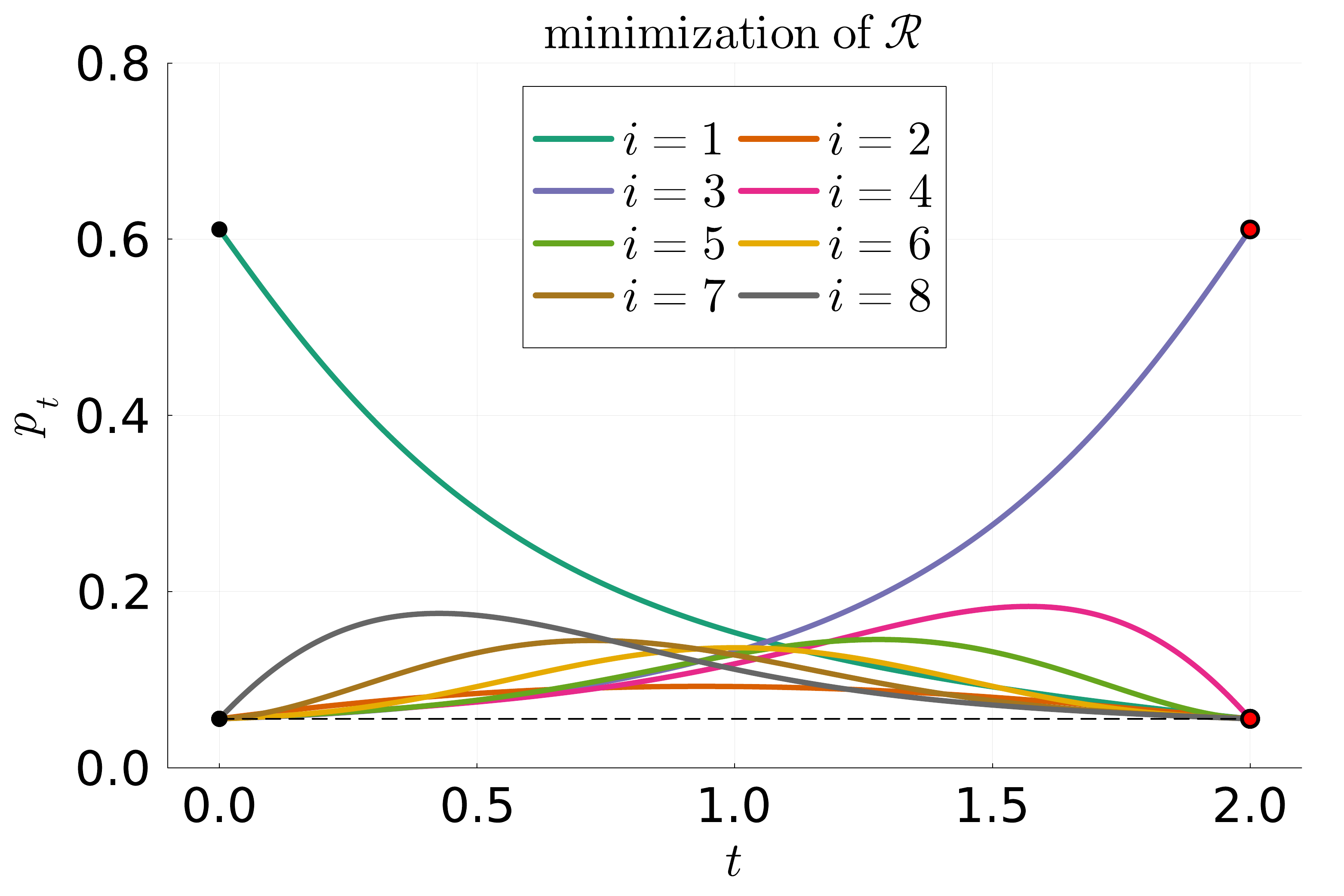}
    \caption{Results for the optimal protocol realizing the transport process illustrated in Fig.~\ref{fig-setup} with $N = 8$, $\omega_{0} = 1$, $\tilde{\omega} = 0.003$, $p_0(1) = 10 p_0(i)$ for all $i \neq 1$, $p_\tau(3) = 10 p_\tau(i)$ for all $i \neq 3$ and $\tau = 2$. The top panel shows the optimal time-evolution of the probability distribution minimizing the entropy production, while the bottom panel shows the one minimizing the renormalized entropy production.}
    \label{fig-protocol}
\end{figure}

\begin{figure}
    \includegraphics[width=0.48\textwidth]{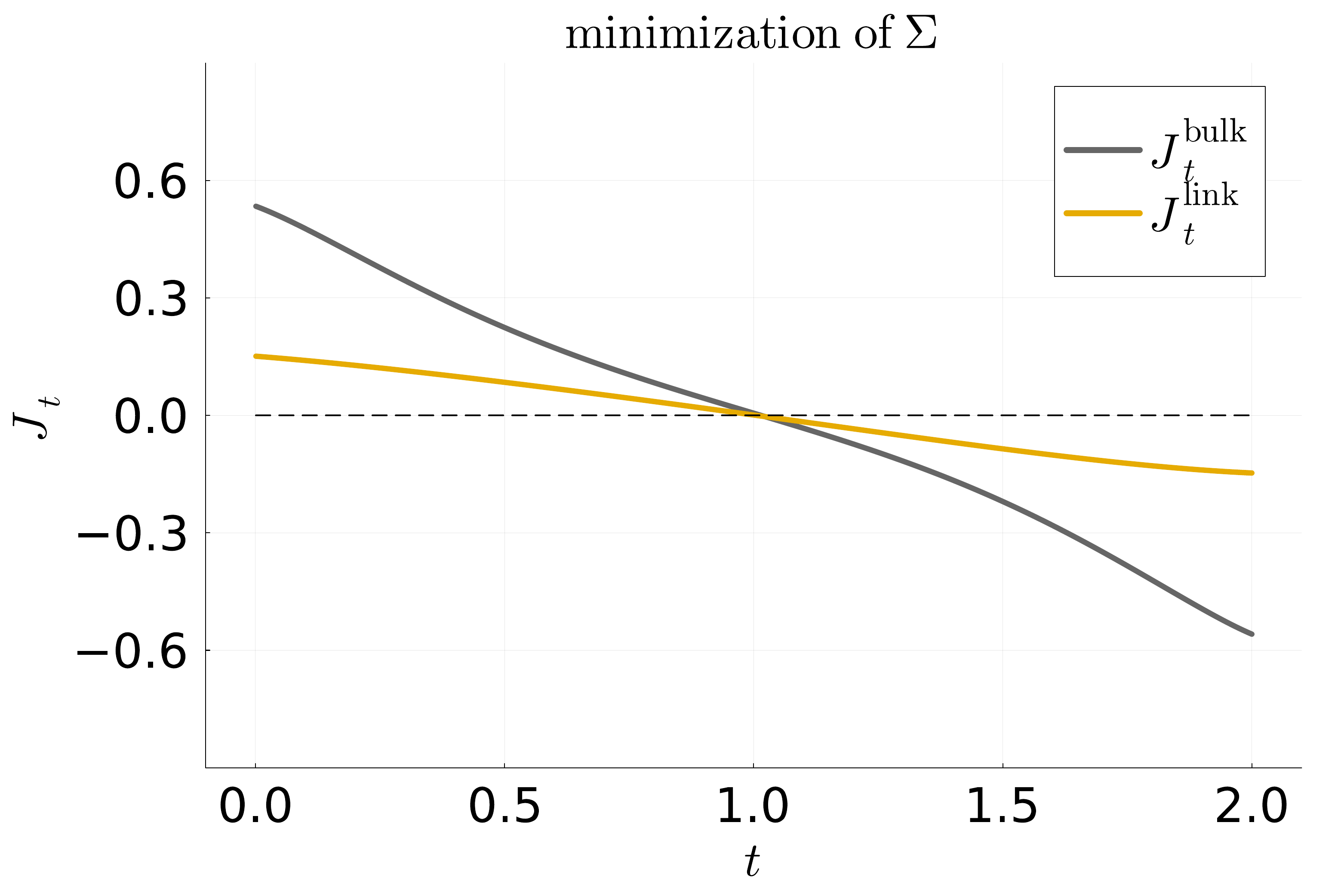}\\
    \includegraphics[width=0.48\textwidth]{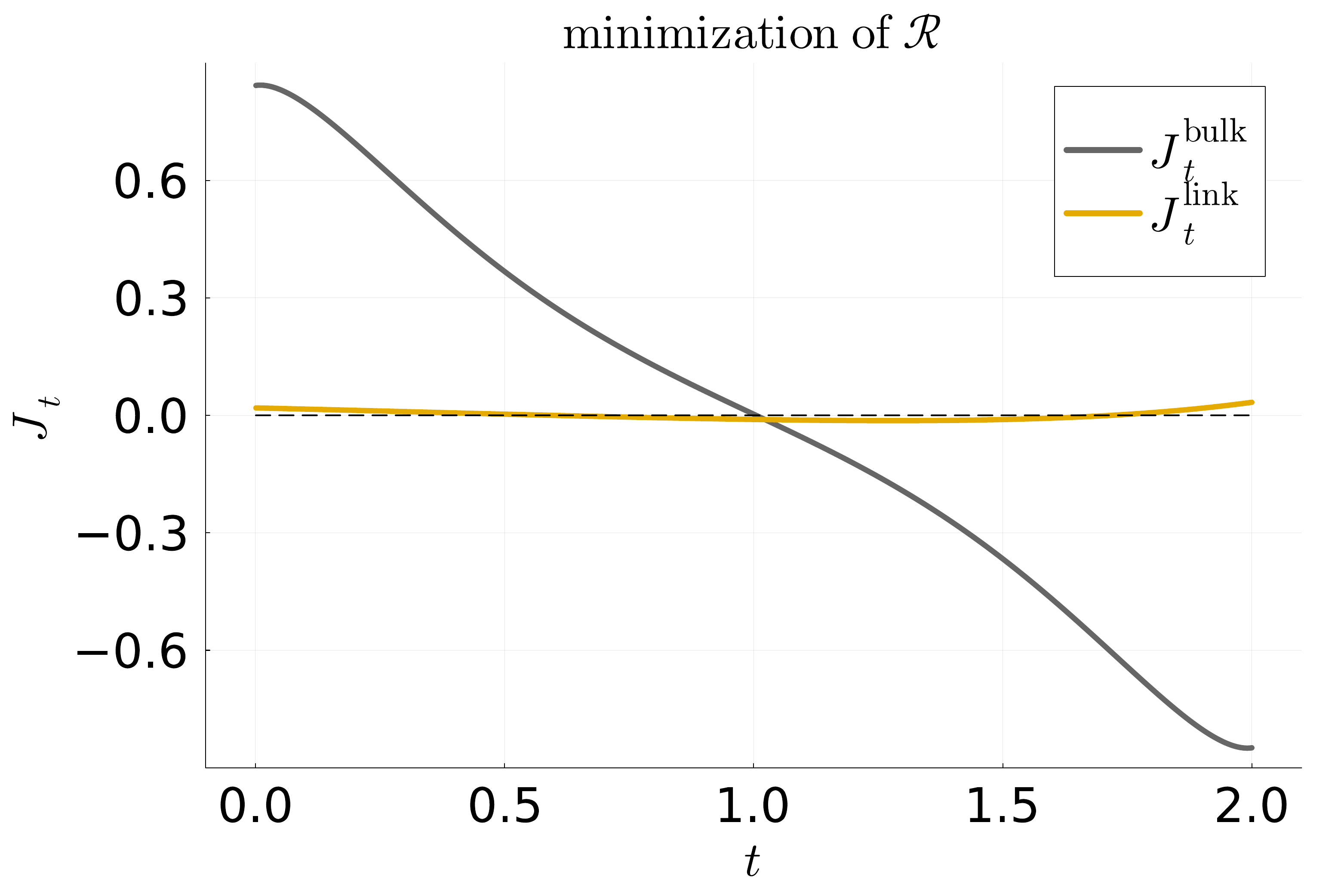}
    \caption{Probability currents corresponding to Fig.~\ref{fig-protocol}. The current through the bulk sites is defined as the difference of the probability current entering site 8 and the probability current exiting site 4, $J_t^\text{bulk} = J_t(8,1) - J_t(3,4)$, while the corresponding current through the link is $J_t^\text{link} = J_t(2,1) - J_t(3,2)$. The top panel shows the probability flows for the protocol minimizing the entropy production, while the bottom panel shows the one minimizing the renormalized entropy production.}
    \label{fig-current}
\end{figure}

\begin{figure}
    \includegraphics[width=0.48\textwidth]{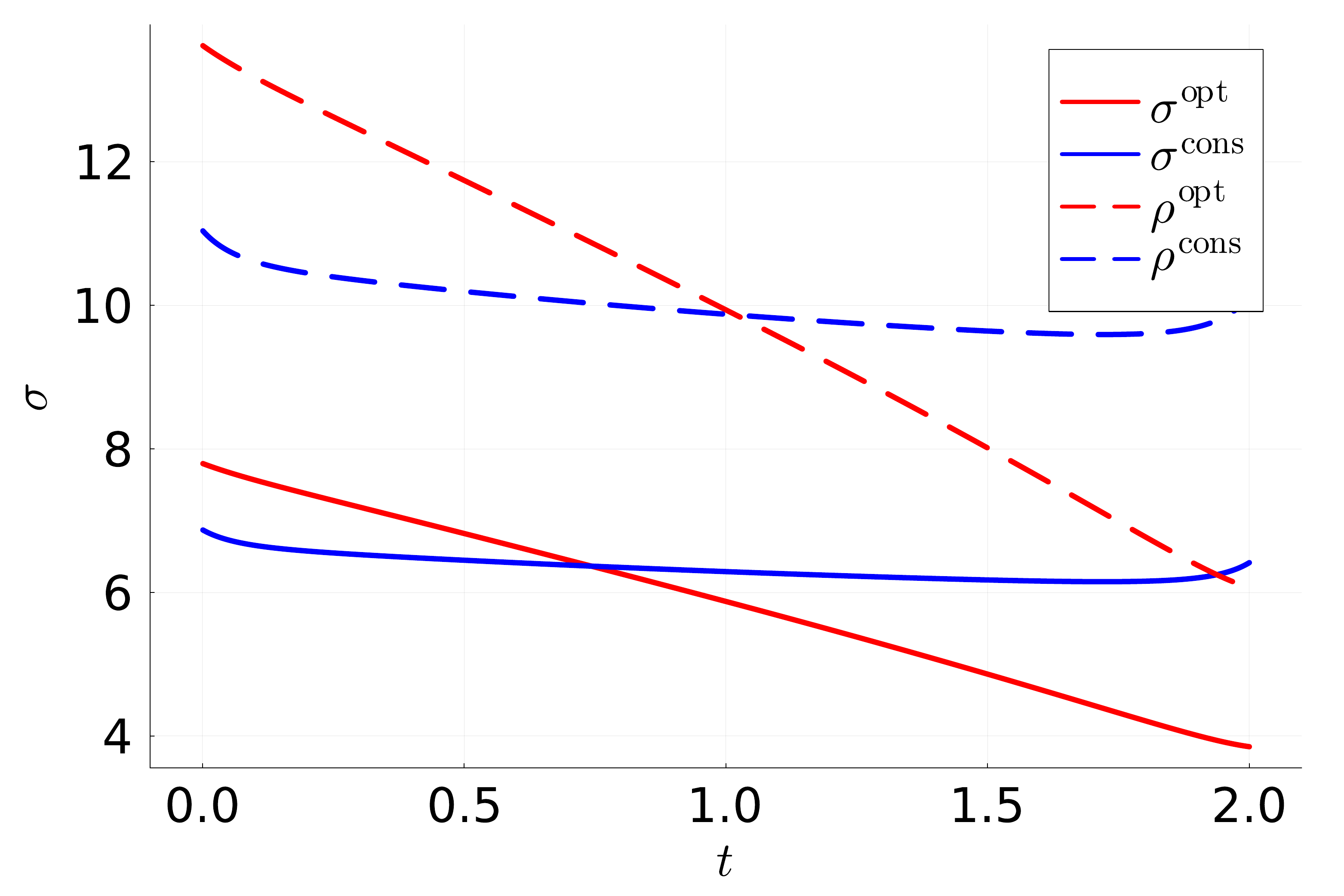}\\
    \includegraphics[width=0.48\textwidth]{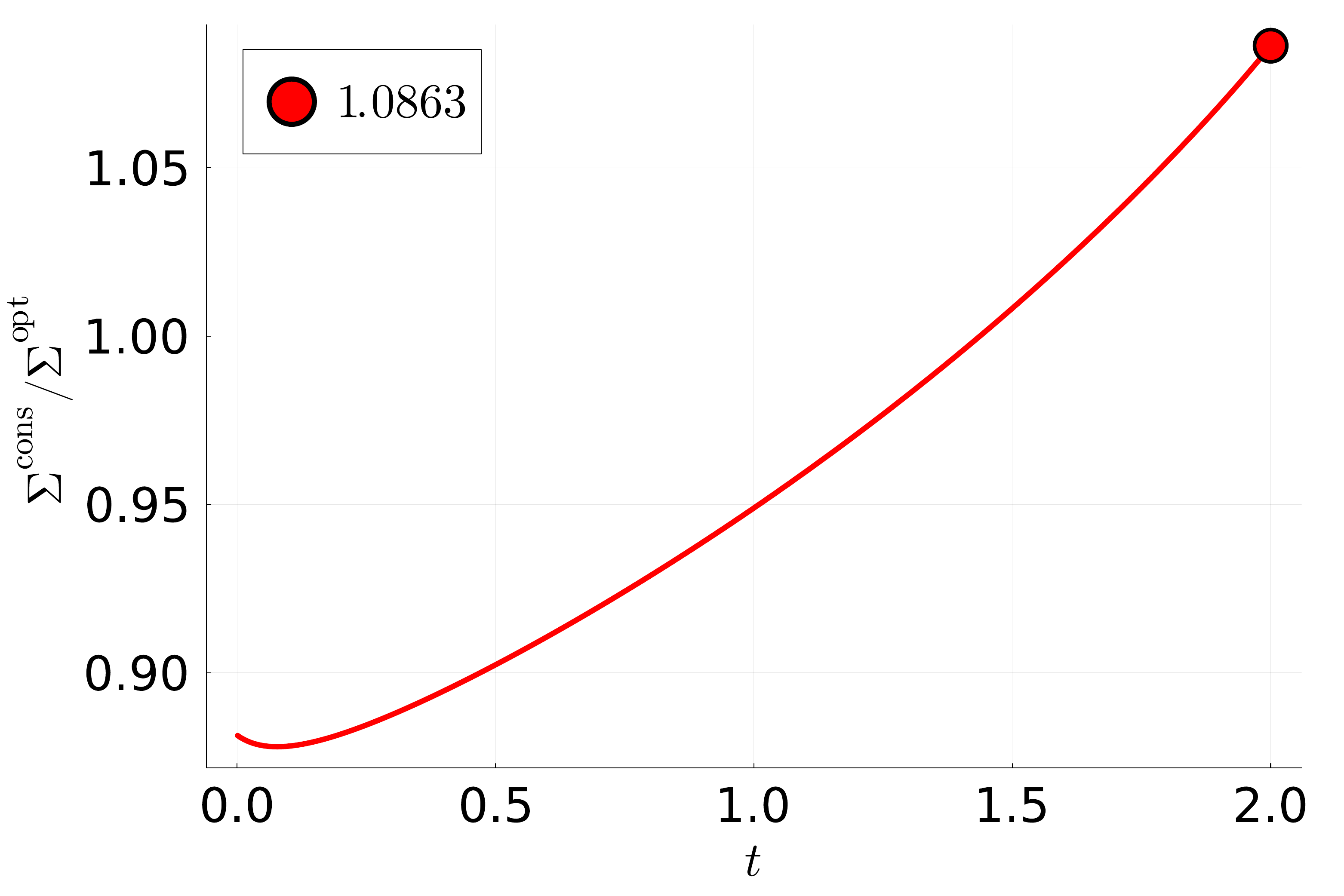}
    \caption{Entropy production and renormalized entropy production for different optimal protocols.
    The top panel depicts the (renormalized) entropy production rate along the non-conservative optimal protocol that minimizes the entropy production (opt) or the conservative protocol that minimzies the renormalized entropy production (cons). The bottom panel shows the ratio of the entropy production for the conservative forces minimizing the renormalized one and the minimal entropy production.}
    \label{fig-entropy}
\end{figure}

\begin{figure}
    \includegraphics[width=0.48\textwidth]{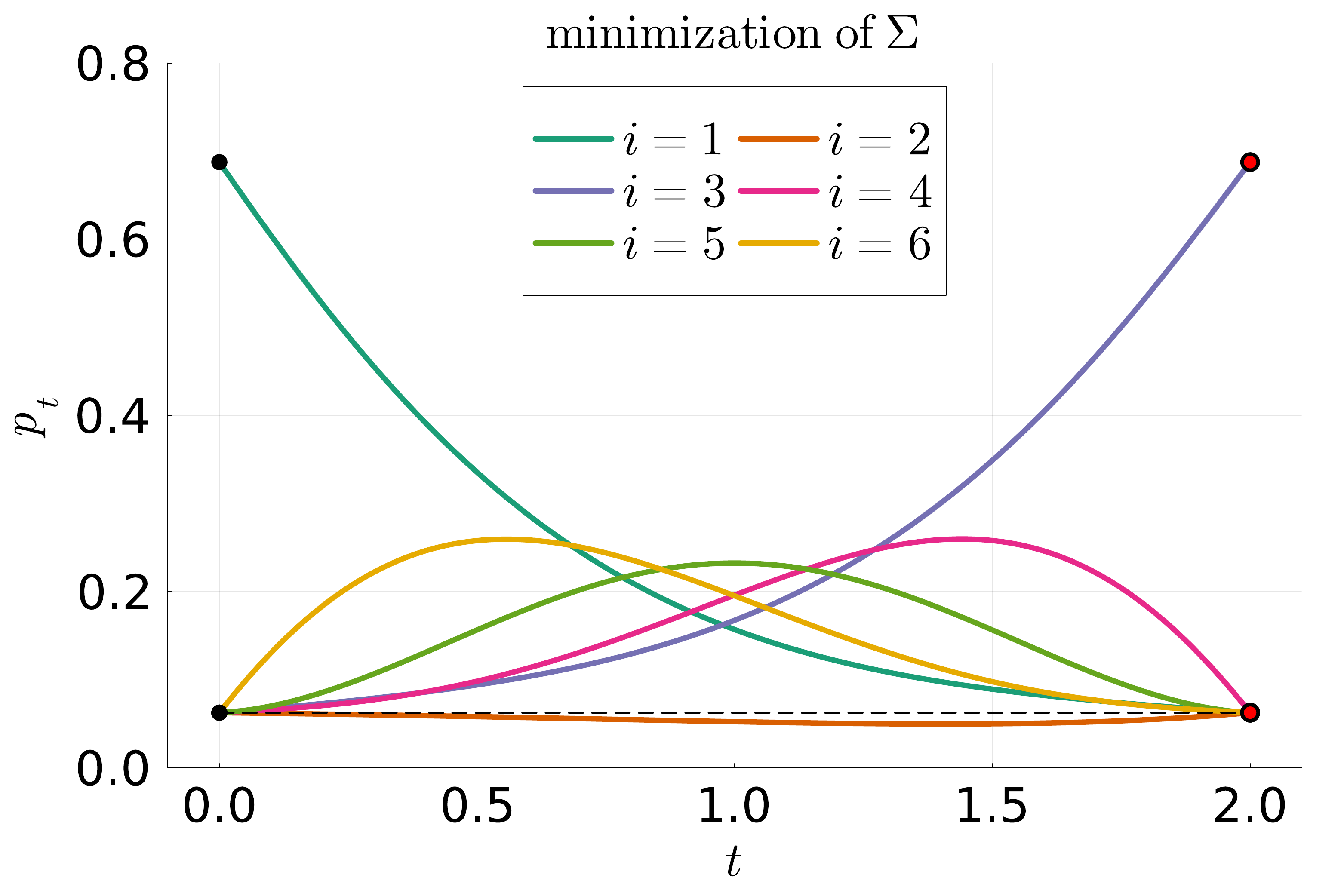}
    \caption{Results for the optimal protocol minimizing entropy production for $N = 6$ sites. Other parameters are as specified in Fig.~\ref{fig-protocol}.}
    \label{fig-protocol-6}
\end{figure}

As a concrete example, we consider the transport process illustrated in Fig.~\ref{fig-setup}.
Eight sites are arranged in a ring configuration and, initially, the system is most likely to be found at site 1, with equal probabilities at all other sites. 
Concretely, the probability to be at site 1 is ten times as large as for the other sites, that is, for eight sites, $p_0(1) = 10/17$ and $p_0(i) = 1/17$ for $i \neq 1$.
Our goal is to drive the transitions between the sites in such a way that, after time $\tau$, the most likely site is site 3, again with $p_\tau(3) = 10 p_\tau(i)$ for $i \neq 3$.
We assume the symmetric part of the transition rates between neighboring sites to be uniform with $\omega(i,i+1) = \omega_0$, except for the transition between sites $1$ and $2$, which occurs at the considerably lower rate $\omega(1,2) = \tilde{\omega} \ll \omega_0$; any other transition rates are set to zero.
A similar setup was considered in Ref.~\cite{shortpaper} and was shown to lead to a quantitative advantage for non-conservative driving protocols.

Since we effectively need to transport probability from site $1$ to site $3$, we can do so along two transition pathways:
Either, we transport probability along the long path through the bulk of states, with relatively fast transitions between each pair of states.
Or, we transport probability along the short path trough site $2$, which, however, requires transport over the much slower transition between sites $1$ and $2$, which effectively acts as a weak link.
In contrast to the short-time optimization, minimizing the (renormalized) entropy production along the process requires us not only to determine the optimal way to distribute the transport along the two pathways, but also along the given time interval.

In Fig.~\ref{fig-protocol} we show the time-evolution of the occupation probabilities between the initial and final state that minimize the entropy production (top) or renormalized entropy production (bottom).
We immediately notice that respective optimal protocols lead to a qualitatively different transport process.
The non-conservative protocol that minimizes the entropy production mostly transports probability through the slow transition between sites 1 and 2, leading the probability of being in site 2 to first increase as probability is transported from site 1 and then decrease as it is transported to site 3, while the occupation probabilities in the bulk remain almost constant.
By contrast, the conservative protocol minimizing the renormalized entropy production transports probability mostly through the bulk, leaving the occupation probability of state $2$ almost unaffected.
This behavior is corroborated by the probability flows through the bulk and the link shown in Fig.~\ref{fig-current}.

Fig.~\ref{fig-entropy} shows the resulting values for the entropy production and renormalized entropy production along the respective protocols.
We note that, even though the non-conservative protocol minimizes the overall entropy production, the corresponding entropy production rate (top panel, solid red line) can be transiently larger than for the near-optimal conservative protocol minimizing the renormalized entropy production (solid blue line).
Specifically, the optimal protocol leads to an initially larger entropy production rate, which is compensated during later parts of the protocol to result in an overall entropy production that is smaller by around $9\%$ (bottom panel).
Conversely, the conservative protocol that minimizes the renormalized entropy production rate actually leads to a larger rate of renormalized entropy production during the second half of the protocol.

We also remark on another, somewhat surprising feature of the finite-time optimal protocols shown in Fig.~\ref{fig-protocol-6}, which shows the results for a similar process but on a ring with six sites.
We see that the optimal protocol now causes the probability of site 2 to go below its initial and final value. Thus, the optimal protocol transiently \enquote{borrows} probability from site 2; this effect can be observed for both the optimal and the near-optimal conservative protocol.
Together with the wave-like motion in which probability propagates through the system from state $1$ via states $6, 5, 4$ in the bulk to state $3$, this highlights the generally non-local and non-monotonic nature of finite-time optimal transport processes.

We stress that, even though the present example is an extreme case in which the transition rates along the transition pathways differ by more than two orders of magnitude, and the near-optimal conservative protocol is qualitatively different from the optimal one, the resulting values for the entropy production only differ by a few percent.
This is supported by the findings of Ref.~\cite{shortpaper}, where, despite considerable numerical efforts, the non-conservative protocol was only found to outperform the conservative one by at most $30\%$.
Consequently, the minimization of the renormalized entropy production results in conservative driving protocols that are typically expected to be almost optimal for minimizing entropy production, well below the factor 2 that is guaranteed by the theory.

\subsection{Summary and formulation as speed limits}

A slight reformulation of our results yields an interpretation as a speed limit. It is, however, important to note that in contrast to results of a similar form reported in, e.g., Ref. \cite{dech22c, vu23a}, the results do not include a Wasserstein distance term.

We define $\bar{\sigma} = \Sigma/\tau$ as the mean entropy production rate over the course of an arbitrary protocol that transforms a given initial configuration $p_0(i) = p_\text{ini}(i)$ into the correct final one $p_\tau(i) = p_\text{fin}(i)$. 
Optimality $\Sigma \geq \Sigma^*$ can be written as
\begin{align}
    \bar{\sigma} \geq \frac{\Av{\lambda_0}_0 - \Av{\lambda_\tau}_\tau}{\tau}
    \label{entropy-speed-limit}
\end{align}
by using \eqref{entropy-finite-time-potential}. This result can be compared to
\begin{align}
    \bar{\rho} \geq \frac{\Av{\lambda_0}_0 - \Av{\lambda_\tau}_\tau}{\tau}
    \label{renentropy-speed-limit}
,\end{align}
which follows from \eqref{renentropy-finite-time-potential} by defining the mean renormalized entropy production $\bar{\rho} = \mathcal{R}/\tau$. In both cases, the average change of the ``potential'' function $\lambda_t(i)$ of the respective optimal solution suffices to determine the implied lower bound. The difference is that only in the latter case, $\lambda_t(i)$ directly characterizes the physical potential landscape via $F_t(i,j) = \lambda_t(j) - \lambda_t(i)$, whereas in the former case we have to refer to the nonlinear relationship
\eqref{entropy-finite-time-force-potential}. 

The relationship between minimal cost and a potential function characterizing the corresponding optimal solution is reminiscent of Kantorovich duality \cite{vill08} and the reduction of the optimal transport problem to the construction of appropriate potential functions \cite{aure12, dech22c}. In particular, one may compare an infinitesimal timestep $\tau = dt$ in \eqref{entropy-speed-limit} to Eq. (38) in Ref. \cite{dech22c}, which characterizes the minimal entropy production rate $\bar{\sigma}^*$ of the optimal protocol under the condition of fixed activity $\mathcal{A}_0$. Transferred to the present notation, this equation reads
\begin{align}
    \bar{\sigma} \geq \bar{\sigma}^* = 2 \Av{d_t \Psi} \, \mathrm{artanh} \left( \frac{\Av{d_t \Psi}}{\mathcal{A}_0} \right)
    \label{entropy-speed-limit-fixed activity}
\end{align}
and involves the rate of change $\Av{d_t \Psi} = (\Av{\Psi}_{dt} - \Av{\Psi}_{0})/dt$ of the potential function $\psi$ that characterizes the optimal solution. Compared to \eqref{entropy-speed-limit}, a crucial difference is that although $\bar{\sigma}$ connects to the potential in a nonlinear manner, the state function $\psi(i)$ nevertheless defines the physical energy landscape without the need of yet another nonlinear transformation like \eqref{entropy-finite-time-force-potential}.

The optimal solution to \eqref{entropy-finite-time-conservative} is hard to characterize directly via a speed limit, because of the additional constraint that the driving forces must be conservative. We can, however, recast \eqref{entropy-finite-time-conservative-inequalities} as
\begin{align}
    \bar{\sigma} \geq \frac{\mathcal{R}^*}{2 \tau} = \frac{\Av{\lambda_0}_0 - \Av{\lambda_\tau}_\tau}{2 \tau}
    \label{entropy-speed-limit-2}
\end{align}
using $\bar{\sigma} = \Sigma/\tau \geq \Sigma^*/\tau$. The inequality holds true for the average entropy production $\bar{\sigma}$ of protocols irrespective of whether they are realized via conservative forces only. In this sense, the solution to \eqref{renentropy-finite-time}, which is simpler to obtain than the solution to \eqref{entropy-finite-time-conservative} but nevertheless incorporates the correct constraints, can be used to formulate a speed limit although it is not the optimal solution.

The different optimization problems that have been discussed in this section are summarized in Table \ref{tab:optimal-summary}.

\begin{table}[ht] 
\caption{\label{tab:optimal-summary} Different optimization problems in Section \ref{sec-finite-time}.}
\begin{ruledtabular}
\begin{tabular}{lccc}
 Optimization problem & \eqref{entropy-finite-time} & \eqref{renentropy-finite-time} & \eqref{entropy-finite-time-conservative} \\
 \colrule 
 Minimized quantity & $\Sigma$ & $\mathcal{R}$ & $\Sigma$ \\
 Constraints & \eqref{master-param}  & \eqref{master-param}  & \eqref{master-param} and \\
 & & & $F_t(i,j)$ conserv. \\
 Sol. conservative? & No & Yes & Yes \\
 Speed limit & (\ref{entropy-speed-limit}), (\ref{entropy-speed-limit-2}) & (\ref{renentropy-speed-limit}) & (\ref{entropy-speed-limit}), (\ref{entropy-speed-limit-2}) \\
\end{tabular}
\end{ruledtabular}
\end{table}

\section{Renormalized entropy and path probability} \label{sec-renorm-path}
While both the entropy production rate $\sigma_t$ and the renormalized entropy production rate $\rho_t$ measure the distance from equilibrium, the former quantity has an intuitive interpretation in terms of the path probabilities of the forward and time-reversed process, \eqref{entropy-diffusion}. Here, establish a similar interpretation for the renormalized entropy production.

\subsection{Renormalized entropy production as a Kullback-Leibler divergence}

We consider a modified dynamics with the exponentially tilted rates
\begin{align}
W_t^Y(i,j) = e^{Y_t(i,j)} W_t(i,j) \label{tilted-rates},
\end{align}
where $Y_t(i,j) = -Y_t(j,i)$ is an arbitrary skew-symmetric matrix.
We stress that all transition rates obtained by this transformation have the same symmetric part,
\begin{align}
\sqrt{W_t(i,j) W_t(j,i)} = \omega_t(i,j) = \sqrt{W_t^Y(i,j) W_t^Y(j,i)} .
\end{align}
Indeed, the tilting corresponds to an additive change in the thermodynamic forces, $F^Y_t(i,j) = F_t(i,j) + 2 Y_t(i,j)$, so all transition rates sharing the same symmetric part can be obtained by a suitable choice of $Y_t(i,j)$.
The path probability associated with these dynamics is denoted $\mathbb{P}^Y(\Gamma)$ and its time-reverse $\mathbb{P}^{Y,\dagger}(\Gamma)$ (with the rates $W_t^{Y,\dagger}(i,j) = W_{\tau-t}^Y(i,j)$).
\begin{widetext}
We can evaluate the Kullback-Leibler (KL) divergences (see Appendix \ref{app-path})
\begin{subequations}
\begin{align}
&D_\text{KL}\big(\mathbb{P}(\Gamma) \Vert \mathbb{P}^Y(\Gamma) \big) = \int_0^\tau dt \sum_{i,j = 1}^N \Omega_t(i,j) \Bigg( \cosh\bigg(\frac{F_t(i,j) + 2 Y_t(i,j)}{2} \bigg) - \cosh\bigg(\frac{F_t(i,j)}{2} \bigg) - Y_t(i,j) \sinh \bigg(\frac{F_t(i,j)}{2} \bigg) \Bigg), \label{dkl-1} \\
&D_\text{KL}\big(\mathbb{P}(\Gamma) \Vert \mathbb{P}^{Y,\dagger}(\Gamma^\dagger) \big) = \int_0^\tau dt \sum_{i,j = 1}^N \Omega_t(i,j) \Bigg( \cosh\bigg(\frac{F_t(i,j) + 2 Y_t(i,j)}{2} \bigg) - \cosh\bigg(\frac{F_t(i,j)}{2} \bigg) + Y_t(i,j) \sinh \bigg(\frac{F_t(i,j)}{2} \bigg) \label{dkl-2} \nn
&\hspace{4cm} + F_t(i,j) \sinh \bigg(\frac{F_t(i,j)}{2} \bigg) \Bigg)  .
\end{align}  \label{dkl}%
\end{subequations}
We note the inequalities $\cosh(x+y) - \cosh(x) - y \sinh(x) \geq 0$ and $\cosh(x+y) - \cosh(x) + y \sinh(x) + 2 x \sinh(x) \geq 0$ for arbitrary $x,y \in \mathbb{R}$, so that each term under the sum is positive. The first inequality can be proved from Lagrange's mean value theorem and monotonicity of $\sinh$, whereas the second inequality can be reduced to the first one, e.g., by setting $y = -t-2x$.
\end{widetext}
We further note the symmetry $D_\text{KL}\big(\mathbb{P}(\Gamma) \Vert \mathbb{P}^{Y,\dagger}(\Gamma^\dagger) \big) = D_\text{KL}\big(\mathbb{P}(\Gamma) \Vert \mathbb{P}^{-Y-F}(\Gamma) \big)$, that is, the time reversal in the second line is equivalent to replacing $Y_t(i,j)$ by $-Y_t(i,j)-F_t(i,j)$.
For $Y_t(i,j) = 0$, the first expression vanishes, while the second expression recovers the entropy production \eqref{entropy-diffusion}.
However, we also have the equivalent expression for the latter,
\begin{align}
\mathcal{S} = D_\text{KL}\big(\mathbb{P}(\Gamma) \Vert \mathbb{P}^{-F}(\Gamma) \big) .
\end{align}
This means that, instead of considering the time-reversed path probability of the original dynamics, we can also characterize entropy production by a dynamics with the transition rates
\begin{align}
W^{-F}_t(i,j) = \omega_t(i,j) \exp \bigg( \frac{\beta}{2} A_t(i,j) \bigg) \frac{p_t(i)}{p_t(j)} \label{entropy-rates},
\end{align}
in which the direction of the forces is reversed.
These rates satisfy the relation
\begin{align}
W^{-F}_t(i,j) p_t(j) &- W^{-F}_t(j,i) p_t(i) \\
& = -\big(W_t(i,j) p_t(j) - W_t(j,i) p_t(i)\big) \n,
\end{align}
that is, they correspond to a reversal of the probability currents.
Choosing $Y_t(i,j) = -F_t(i,j)/2$, we obtain
\begin{align}
D_\text{KL}\big(&\mathbb{P}(\Gamma) \Vert \mathbb{P}^{-F/2}(\Gamma) \big) = D_\text{KL}\big(\mathbb{P}(\Gamma) \Vert \mathbb{P}^{-F/2,\dagger}(\Gamma^\dagger) \big) \label{eta-path} =  \frac{1}{4} \mathcal{R} .
\end{align}
Thus, up to a factor $1/4$, the renormalized entropy production can be written as a KL divergence between the original dynamics and a dynamics with the rates
\begin{align}
W^{-F/2}_t(i,j) = \omega_t(i,j) \exp \bigg( \frac{1}{2} \big( \ln p_t(i) - \ln p_t(j) \big) \bigg) \label{eta-rates} .
\end{align}
These rates satisfy the detailed balance condition \eqref{detailed-balance} with the time-dependent state of the original dynamics, that is, for these rates, the state $p_t(i)$ is  the instantaneous equilibrium state,
\begin{align}
W^{-F/2}_t(i,j)p_t(j) - W_t^{-F/2}(j,i) p_t(i) = 0.
\end{align}
Intuitively, while the entropy production $\mathcal{S}$ quantifies the difference between the original dynamics and a fictitious dynamics with the same state but reversed probability currents, the renormalized entropy production $\mathcal{R}$ quantifies the difference between the original dynamics and a dynamics with the same state but vanishing probability currents.
This interpretation allows both quantities to be understood as a measure of irreversibility, the difference being whether we compare between the original and the \textit{reversed} dynamics, or between the original and the \textit{reversible} dynamics with the same state.
We remark that in the continuous case, the latter corresponds to a force $\tilde{\bm{F}}_t(\bm{x}) = \bm{F}_t(\bm{x}) - \bm{\nu}_t(\bm{x})/\mu = T \grad \ln p_t(\bm{x})$.
Computing the KL divergence with the corresponding path probability $\mathbb{P}^{-\nu}(\Gamma)$, we find
\begin{align}
D_\text{KL}\big(&\mathbb{P}(\Gamma) \Vert \mathbb{P}^{-\nu}(\Gamma) \big) = D_\text{KL}\big( \mathbb{P}(\Gamma) \Vert \mathbb{P}^{-\nu,\dagger}(\Gamma^\dagger) \big) = \frac{1}{4} \mathcal{S} ,
\end{align}
which realizes the lower bound $\rho_t = \sigma_t$ in \eqref{eta-sigma-inequality}.
Thus, the definition of the renormalized entropy production via the path ratio \eqref{eta-path} is consistent between the diffusion and jump case.


\subsection{Renormalized entropy production along a trajectory and a fluctuation relation}

In the above discussion, we only considered the KL divergence between different path probabilities, which already involves an average over trajectories.
However, we may also consider the path probability for a given trajectory.
We introduce a reference process with transition rates $W_t^0(i,j)$ and initial state $p_0^0(i)$.
The path probability of a given trajectory can then be expressed as
\begin{align}
&\frac{\mathbb{P}(\Gamma)}{\mathbb{P}^0(\Gamma)} = \exp \Bigg( \int_0^\tau dt \sum_{k=1}^N \big( W_t(k,i(t))^0 - W_t(k,i(t)) \big) \nn
&\qquad + \sum_{\text{jumps}} \ln \bigg( \frac{W_t(i(t+dt),i(t))}{W_t^0(i(t+dt),i(t))} \bigg) + \ln \bigg(\frac{p_0(i(0))}{p_0^0(i(0))} \bigg) \Bigg) \label{path-trajectory} ,
\end{align}
where $\mathbb{P}^0(\Gamma)$ is the path probability of the reference process and $i(t)$ denotes the instantaneous state of the system at time $t$.
Here, we sum over all jumps along a given trajectory, that is, all instances in time where $i(t+dt)$ differs from $i(t)$.
If we choose the reference process to have uniform rates $W_t^0(i,j) = \omega^0$ and occupation probability $p_0^0(i) = 1/N$, then this can be written as
\begin{align}
\frac{\mathbb{P}(\Gamma)}{\mathbb{P}^0(\Gamma)} = \exp \bigg( \omega_0 \tau - \mathcal{D}(\Gamma) + \frac{1}{2} \widetilde{\mathcal{Q}}(\Gamma) \bigg) N p_0(i(0)),
\end{align}
where $\mathcal{D}(\Gamma)$ is a quantity called frenesy \cite{maes2020a}, which corresponds to the time-symmetric part of the path probability and is given by
\begin{align}
\mathcal{D}(\Gamma) = \int_0^\tau dt &\sum_{k=1}^N W_t(k,i(t)) \\
& - \sum_{\text{jumps}} \ln \bigg( \frac{\omega_t(i(t+dt),i(t)}{\omega_0} \bigg) \n,
\end{align}
whereas $\widetilde{\mathcal{Q}}(\Gamma)$ can be interpreted as the dissipated heat along the trajectory,
\begin{align}
\widetilde{\mathcal{Q}}(\Gamma) &= \sum_{\text{jumps}} \ln \bigg(\frac{W_t(i(t+dt),i(t))}{W_t(i(t),i(t+dt))} \bigg) \\
& = -\beta \sum_{\text{jumps}} A_t(i(t+dt),i(t)) \n ,
\end{align}
which corresponds to the time-antisymmetric part of the path probability.
More precisely, $\mathcal{D}(\Gamma)$ measures the excess in frenesy, that is, the time-symmetric activity in the system, compared to a reference process.
If, instead of considering a fixed reference process, we take the latter to be a modified version of the original process, then \eqref{path-trajectory} provides the definition of various quantities on a trajectory level.
For example, choosing the \enquote{time-reversed} rates \eqref{entropy-rates}, we obtain
\begin{align}
\mathcal{S}(\Gamma) &= \ln \bigg(\frac{\mathbb{P}(\Gamma)}{\mathbb{P}^{-F}(\Gamma)} \bigg) = \int_0^\tau dt \ d_t \ln p_t(i(t)) \\
&\qquad + \sum_{\text{jumps}} \ln \bigg( \frac{W_t(i(t+dt),i(t)) p_t(i(t))}{W_t(i(t),i(t+dt)) p_t(i(t+dt))} \bigg) \n ,
\end{align}
which can be interpreted as a stochastic entropy production along the trajectory in the sense that its average is equal to $\mathcal{S}$, \eqref{entropy-diffusion}.
On the other hand, choosing the \enquote{detailed-balanced} rates \eqref{eta-rates}, we obtain
\begin{align}
\frac{1}{4} \mathcal{R}(\Gamma) &= \ln \bigg(\frac{\mathbb{P}(\Gamma)}{\mathbb{P}^{-F/2}(\Gamma)} \bigg) \\
& = \int_0^\tau dt \sum_{k=1}^N \bigg( \omega_t(k,i(t)) \sqrt{\frac{p_t(k)}{p_t(i(k))}} - W_t(k,i(t)) \bigg) \nn
&\quad + \frac{1}{2} \sum_{\text{jumps}} \ln \bigg( \frac{W_t(i(t+dt),i(t)) p_t(i(t))}{W_t(i(t),i(t+dt)) p_t(i(t+dt))} \bigg) \n ,
\end{align}
which averages to $\mathcal{R}$, \eqref{renorm-entropy}, and can thus be interpreted as a stochastic version of the renormalized entropy production rate.
Note that from the above expressions, we see that the stochastic entropy production as well as its renormalized version each satisfy an integral fluctuation theorem,
\begin{subequations}
\begin{align}
\Av{e^{-\mathcal{S}}} &= \int d\Gamma \ e^{-\mathcal{S}(\Gamma)} \mathbb{P}(\Gamma) = 1, \\
\Av{e^{-\frac{1}{4}\mathcal{R}}} &= \int d\Gamma \ e^{-\frac{1}{4}\mathcal{R}(\Gamma)} \mathbb{P}(\Gamma) = 1 .
\end{align}
\end{subequations}
This follows from the fact that $\mathbb{P}^{-F}(\Gamma)$ and $\mathbb{P}^{-F/2}(\Gamma)$ are likewise normalized path probabilities.

\section{Discussion and outlook} \label{sec-discussion}

A fundamental distinction between jump processes and their counterparts in continuous space is that concepts held separate in the former case can collapse and coincide for overdamped diffusion processes. In the context of thermodynamic uncertainty relations, this applies to entropy production and pseudo-entropy production \cite{shir21}. For optimal transport, we have introduced renormalized entropy production as a measure of distance to equilibrium whose minimization for finite-time processes results in a protocol that can be realized via conservative forces. Thus, it may be fair to interpret this quantity as a ``distance to conservativity''.

Since renormalized entropy production approaches the usual entropy production near equilibrium, significant differences between these two optimization problems can only become apparent for systems with a nontrivial topology under far-from-equilibrium conditions. From a biophysical perspective, this situation is highly relevant and applies to, e.g., molecular motors (Kolo, etc). It is therefore promising albeit speculative to study the role of conservative and nonconservative forces in such systems. For example, thermodynamic and energetic constraints in such systems may result in optimization strategies that do not purely aim at minimizing dissipation but balance this goal with, e.g., realizing this goal with conservative forces.

From a theoretical point of view, it is also worthwhile to discuss different parametrizations for the transition rates and therefore different optimization problems. In which cases can we establish analogues of the renormalized entropy production that result in optimal solutions realizable with conservative forces? This approach offers a way to classify optimization problems based on whether they are inherently nonconservative. If an analogue of the renormalized entropy production can be established, it is worthwhile to establish the properties of these quantities. It would be particularly interesting to compare the different levels of structure that different optimal transport problems have: For example, under the given parametrization neither the minimization of entropy production nor the minimzation of renormalized entropy production yields the appealing Wasserstein structure present in continuous systems \cite{aure11} or under the assumption of constant state mobility \cite{vu23a} or total activity \cite{dech22c}. Nevertheless, for both minimization problems the optimal solution is characterized via a ``potential'' function, a structural property reminiscent of Kantorovich-Rubinstein duality \cite{vill08} that is lost if we consider, e.g., optimization of entropy production under the additional constraint of a conseravtive protocol.



\begin{thebibliography}{41}%
\makeatletter
\providecommand \@ifxundefined [1]{%
 \@ifx{#1\undefined}
}%
\providecommand \@ifnum [1]{%
 \ifnum #1\expandafter \@firstoftwo
 \else \expandafter \@secondoftwo
 \fi
}%
\providecommand \@ifx [1]{%
 \ifx #1\expandafter \@firstoftwo
 \else \expandafter \@secondoftwo
 \fi
}%
\providecommand \natexlab [1]{#1}%
\providecommand \enquote  [1]{``#1''}%
\providecommand \bibnamefont  [1]{#1}%
\providecommand \bibfnamefont [1]{#1}%
\providecommand \citenamefont [1]{#1}%
\providecommand \href@noop [0]{\@secondoftwo}%
\providecommand \href [0]{\begingroup \@sanitize@url \@href}%
\providecommand \@href[1]{\@@startlink{#1}\@@href}%
\providecommand \@@href[1]{\endgroup#1\@@endlink}%
\providecommand \@sanitize@url [0]{\catcode `\\12\catcode `\$12\catcode `\&12\catcode `\#12\catcode `\^12\catcode `\_12\catcode `\%12\relax}%
\providecommand \@@startlink[1]{}%
\providecommand \@@endlink[0]{}%
\providecommand \url  [0]{\begingroup\@sanitize@url \@url }%
\providecommand \@url [1]{\endgroup\@href {#1}{\urlprefix }}%
\providecommand \urlprefix  [0]{URL }%
\providecommand \Eprint [0]{\href }%
\providecommand \doibase [0]{https://doi.org/}%
\providecommand \selectlanguage [0]{\@gobble}%
\providecommand \bibinfo  [0]{\@secondoftwo}%
\providecommand \bibfield  [0]{\@secondoftwo}%
\providecommand \translation [1]{[#1]}%
\providecommand \BibitemOpen [0]{}%
\providecommand \bibitemStop [0]{}%
\providecommand \bibitemNoStop [0]{.\EOS\space}%
\providecommand \EOS [0]{\spacefactor3000\relax}%
\providecommand \BibitemShut  [1]{\csname bibitem#1\endcsname}%
\let\auto@bib@innerbib\@empty
\bibitem [{\citenamefont {Sekimoto}(2010)}]{seki10}%
  \BibitemOpen
  \bibfield  {author} {\bibinfo {author} {\bibfnamefont {K.}~\bibnamefont {Sekimoto}},\ }\href {https://doi.org/10.1007/978-3-642-05411-2} {\emph {\bibinfo {title} {Stochastic {{Energetics}}}}},\ \bibinfo {series} {Lecture {{Notes}} in {{Physics}}}, Vol.\ \bibinfo {volume} {799}\ (\bibinfo  {publisher} {Springer Berlin Heidelberg},\ \bibinfo {address} {Berlin, Heidelberg},\ \bibinfo {year} {2010})\BibitemShut {NoStop}%
\bibitem [{\citenamefont {Peliti}\ and\ \citenamefont {Pigolotti}(2021)}]{peli21}%
  \BibitemOpen
  \bibfield  {author} {\bibinfo {author} {\bibfnamefont {L.}~\bibnamefont {Peliti}}\ and\ \bibinfo {author} {\bibfnamefont {S.}~\bibnamefont {Pigolotti}},\ }\href@noop {} {\emph {\bibinfo {title} {Stochastic {{Thermodynamics}}: {{An Introduction}}}}}\ (\bibinfo  {publisher} {Princeton University Press},\ \bibinfo {year} {2021})\BibitemShut {NoStop}%
\bibitem [{\citenamefont {Shiraishi}(2023)}]{shir23}%
  \BibitemOpen
  \bibfield  {author} {\bibinfo {author} {\bibfnamefont {N.}~\bibnamefont {Shiraishi}},\ }\href {https://doi.org/10.1007/978-981-19-8186-9} {\emph {\bibinfo {title} {An {{Introduction}} to {{Stochastic Thermodynamics}}: {{From Basic}} to {{Advanced}}}}},\ \bibinfo {series} {Fundamental {{Theories}} of {{Physics}}}, Vol.\ \bibinfo {volume} {212}\ (\bibinfo  {publisher} {Springer Nature Singapore},\ \bibinfo {address} {Singapore},\ \bibinfo {year} {2023})\BibitemShut {NoStop}%
\bibitem [{\citenamefont {Seifert}(2025)}]{seif25}%
  \BibitemOpen
  \bibfield  {author} {\bibinfo {author} {\bibfnamefont {U.}~\bibnamefont {Seifert}},\ }\href@noop {} {\emph {\bibinfo {title} {Stochastic thermodynamics}}}\ (\bibinfo  {publisher} {Cambridge University Press},\ \bibinfo {address} {Cambridge, England},\ \bibinfo {year} {2025})\BibitemShut {NoStop}%
\bibitem [{\citenamefont {Schmiedl}\ and\ \citenamefont {Seifert}(2007)}]{schm07a}%
  \BibitemOpen
  \bibfield  {author} {\bibinfo {author} {\bibfnamefont {T.}~\bibnamefont {Schmiedl}}\ and\ \bibinfo {author} {\bibfnamefont {U.}~\bibnamefont {Seifert}},\ }\bibfield  {title} {\bibinfo {title} {Optimal {{Finite-Time Processes In Stochastic Thermodynamics}}},\ }\href {https://doi.org/10.1103/PhysRevLett.98.108301} {\bibfield  {journal} {\bibinfo  {journal} {Physical Review Letters}\ }\textbf {\bibinfo {volume} {98}},\ \bibinfo {pages} {108301} (\bibinfo {year} {2007})}\BibitemShut {NoStop}%
\bibitem [{\citenamefont {Monge}(1781)}]{mong81}%
  \BibitemOpen
  \bibfield  {author} {\bibinfo {author} {\bibfnamefont {G.}~\bibnamefont {Monge}},\ }\bibfield  {title} {\bibinfo {title} {Memoire sur la theorie des deblais et des remblais},\ }\href {https://ci.nii.ac.jp/naid/10018386702/en/} {\bibfield  {journal} {\bibinfo  {journal} {Histoire de l'Academie Royale des Sciences de Paris}\ } (\bibinfo {year} {1781})}\BibitemShut {NoStop}%
\bibitem [{\citenamefont {Benamou}\ and\ \citenamefont {Brenier}(2000)}]{bena00}%
  \BibitemOpen
  \bibfield  {author} {\bibinfo {author} {\bibfnamefont {J.-D.}\ \bibnamefont {Benamou}}\ and\ \bibinfo {author} {\bibfnamefont {Y.}~\bibnamefont {Brenier}},\ }\bibfield  {title} {\bibinfo {title} {A computational fluid mechanics solution to the {Monge-Kantorovich} mass transfer problem},\ }\href@noop {} {\bibfield  {journal} {\bibinfo  {journal} {Numer. Math.}\ }\textbf {\bibinfo {volume} {84}},\ \bibinfo {pages} {375} (\bibinfo {year} {2000})}\BibitemShut {NoStop}%
\bibitem [{\citenamefont {Villani}(2008)}]{vill08}%
  \BibitemOpen
  \bibfield  {author} {\bibinfo {author} {\bibfnamefont {C.}~\bibnamefont {Villani}},\ }\href {https://books.google.co.jp/books?id=hV8o5R7\_5tkC} {\emph {\bibinfo {title} {Optimal Transport: Old and New}}},\ Grundlehren der mathematischen Wissenschaften\ (\bibinfo  {publisher} {Springer Berlin Heidelberg},\ \bibinfo {year} {2008})\BibitemShut {NoStop}%
\bibitem [{\citenamefont {Aurell}\ \emph {et~al.}(2011)\citenamefont {Aurell}, \citenamefont {{Mej{\'i}a-Monasterio}},\ and\ \citenamefont {{Muratore-Ginanneschi}}}]{aure11}%
  \BibitemOpen
  \bibfield  {author} {\bibinfo {author} {\bibfnamefont {E.}~\bibnamefont {Aurell}}, \bibinfo {author} {\bibfnamefont {C.}~\bibnamefont {{Mej{\'i}a-Monasterio}}},\ and\ \bibinfo {author} {\bibfnamefont {P.}~\bibnamefont {{Muratore-Ginanneschi}}},\ }\bibfield  {title} {\bibinfo {title} {Optimal {{Protocols}} and {{Optimal Transport}} in {{Stochastic Thermodynamics}}},\ }\href {https://doi.org/10.1103/PhysRevLett.106.250601} {\bibfield  {journal} {\bibinfo  {journal} {Physical Review Letters}\ }\textbf {\bibinfo {volume} {106}},\ \bibinfo {pages} {250601} (\bibinfo {year} {2011})}\BibitemShut {NoStop}%
\bibitem [{\citenamefont {Aurell}\ \emph {et~al.}(2012)\citenamefont {Aurell}, \citenamefont {Gaw{\c e}dzki}, \citenamefont {{Mej{\'i}a-Monasterio}}, \citenamefont {Mohayaee},\ and\ \citenamefont {{Muratore-Ginanneschi}}}]{aure12}%
  \BibitemOpen
  \bibfield  {author} {\bibinfo {author} {\bibfnamefont {E.}~\bibnamefont {Aurell}}, \bibinfo {author} {\bibfnamefont {K.}~\bibnamefont {Gaw{\c e}dzki}}, \bibinfo {author} {\bibfnamefont {C.}~\bibnamefont {{Mej{\'i}a-Monasterio}}}, \bibinfo {author} {\bibfnamefont {R.}~\bibnamefont {Mohayaee}},\ and\ \bibinfo {author} {\bibfnamefont {P.}~\bibnamefont {{Muratore-Ginanneschi}}},\ }\bibfield  {title} {\bibinfo {title} {Refined {{Second Law}} of {{Thermodynamics}} for {{Fast Random Processes}}},\ }\href {https://doi.org/10.1007/s10955-012-0478-x} {\bibfield  {journal} {\bibinfo  {journal} {Journal of Statistical Physics}\ }\textbf {\bibinfo {volume} {147}},\ \bibinfo {pages} {487} (\bibinfo {year} {2012})}\BibitemShut {NoStop}%
\bibitem [{\citenamefont {Nakazato}\ and\ \citenamefont {Ito}(2021)}]{naka21}%
  \BibitemOpen
  \bibfield  {author} {\bibinfo {author} {\bibfnamefont {M.}~\bibnamefont {Nakazato}}\ and\ \bibinfo {author} {\bibfnamefont {S.}~\bibnamefont {Ito}},\ }\bibfield  {title} {\bibinfo {title} {Geometrical aspects of entropy production in stochastic thermodynamics based on {Wasserstein} distance},\ }\href@noop {} {\bibfield  {journal} {\bibinfo  {journal} {Phys. Rev. Research}\ }\textbf {\bibinfo {volume} {3}},\ \bibinfo {pages} {043093} (\bibinfo {year} {2021})}\BibitemShut {NoStop}%
\bibitem [{\citenamefont {B{\'e}rut}\ \emph {et~al.}(2012)\citenamefont {B{\'e}rut}, \citenamefont {Arakelyan}, \citenamefont {Petrosyan}, \citenamefont {Ciliberto}, \citenamefont {Dillenschneider},\ and\ \citenamefont {Lutz}}]{beru12}%
  \BibitemOpen
  \bibfield  {author} {\bibinfo {author} {\bibfnamefont {A.}~\bibnamefont {B{\'e}rut}}, \bibinfo {author} {\bibfnamefont {A.}~\bibnamefont {Arakelyan}}, \bibinfo {author} {\bibfnamefont {A.}~\bibnamefont {Petrosyan}}, \bibinfo {author} {\bibfnamefont {S.}~\bibnamefont {Ciliberto}}, \bibinfo {author} {\bibfnamefont {R.}~\bibnamefont {Dillenschneider}},\ and\ \bibinfo {author} {\bibfnamefont {E.}~\bibnamefont {Lutz}},\ }\bibfield  {title} {\bibinfo {title} {Experimental verification of {{Landauer}}'s principle linking information and thermodynamics},\ }\href {https://doi.org/10.1038/nature10872} {\bibfield  {journal} {\bibinfo  {journal} {Nature}\ }\textbf {\bibinfo {volume} {483}},\ \bibinfo {pages} {187} (\bibinfo {year} {2012})}\BibitemShut {NoStop}%
\bibitem [{\citenamefont {Zulkowski}\ and\ \citenamefont {DeWeese}(2014)}]{zulk14}%
  \BibitemOpen
  \bibfield  {author} {\bibinfo {author} {\bibfnamefont {P.~R.}\ \bibnamefont {Zulkowski}}\ and\ \bibinfo {author} {\bibfnamefont {M.~R.}\ \bibnamefont {DeWeese}},\ }\bibfield  {title} {\bibinfo {title} {Optimal finite-time erasure of a classical bit},\ }\href {https://doi.org/10.1103/PhysRevE.89.052140} {\bibfield  {journal} {\bibinfo  {journal} {Physical Review E}\ }\textbf {\bibinfo {volume} {89}},\ \bibinfo {pages} {052140} (\bibinfo {year} {2014})}\BibitemShut {NoStop}%
\bibitem [{\citenamefont {Dechant}\ and\ \citenamefont {Sakurai}(2019)}]{dech19}%
  \BibitemOpen
  \bibfield  {author} {\bibinfo {author} {\bibfnamefont {A.}~\bibnamefont {Dechant}}\ and\ \bibinfo {author} {\bibfnamefont {Y.}~\bibnamefont {Sakurai}},\ }\href@noop {} {\bibinfo {title} {Thermodynamic interpretation of wasserstein distance}} (\bibinfo {year} {2019}),\ \Eprint {https://arxiv.org/abs/arXiv:1912.08405} {arXiv:1912.08405} \BibitemShut {NoStop}%
\bibitem [{\citenamefont {Proesmans}\ \emph {et~al.}(2020)\citenamefont {Proesmans}, \citenamefont {Ehrich},\ and\ \citenamefont {Bechhoefer}}]{proe20a}%
  \BibitemOpen
  \bibfield  {author} {\bibinfo {author} {\bibfnamefont {K.}~\bibnamefont {Proesmans}}, \bibinfo {author} {\bibfnamefont {J.}~\bibnamefont {Ehrich}},\ and\ \bibinfo {author} {\bibfnamefont {J.}~\bibnamefont {Bechhoefer}},\ }\bibfield  {title} {\bibinfo {title} {Finite-{{Time Landauer Principle}}},\ }\href {https://doi.org/10.1103/PhysRevLett.125.100602} {\bibfield  {journal} {\bibinfo  {journal} {Physical Review Letters}\ }\textbf {\bibinfo {volume} {125}},\ \bibinfo {pages} {100602} (\bibinfo {year} {2020})}\BibitemShut {NoStop}%
\bibitem [{\citenamefont {Blaber}\ and\ \citenamefont {Sivak}(2023)}]{blab23}%
  \BibitemOpen
  \bibfield  {author} {\bibinfo {author} {\bibfnamefont {S.}~\bibnamefont {Blaber}}\ and\ \bibinfo {author} {\bibfnamefont {D.~A.}\ \bibnamefont {Sivak}},\ }\bibfield  {title} {\bibinfo {title} {Optimal control in stochastic thermodynamics},\ }\href {https://doi.org/10.1088/2399-6528/acbf04} {\bibfield  {journal} {\bibinfo  {journal} {Journal of Physics Communications}\ }\textbf {\bibinfo {volume} {7}},\ \bibinfo {pages} {033001} (\bibinfo {year} {2023})}\BibitemShut {NoStop}%
\bibitem [{\citenamefont {Oikawa}\ \emph {et~al.}(2025)\citenamefont {Oikawa}, \citenamefont {Nakayama}, \citenamefont {Ito}, \citenamefont {Sagawa},\ and\ \citenamefont {Toyabe}}]{oika25}%
  \BibitemOpen
  \bibfield  {author} {\bibinfo {author} {\bibfnamefont {S.}~\bibnamefont {Oikawa}}, \bibinfo {author} {\bibfnamefont {Y.}~\bibnamefont {Nakayama}}, \bibinfo {author} {\bibfnamefont {S.}~\bibnamefont {Ito}}, \bibinfo {author} {\bibfnamefont {T.}~\bibnamefont {Sagawa}},\ and\ \bibinfo {author} {\bibfnamefont {S.}~\bibnamefont {Toyabe}},\ }\bibfield  {title} {\bibinfo {title} {Experimentally achieving minimal dissipation via thermodynamically optimal transport},\ }\href {https://doi.org/10.1038/s41467-025-66519-9} {\bibfield  {journal} {\bibinfo  {journal} {Nature Communications}\ }\textbf {\bibinfo {volume} {16}},\ \bibinfo {pages} {10424} (\bibinfo {year} {2025})}\BibitemShut {NoStop}%
\bibitem [{\citenamefont {Barato}\ and\ \citenamefont {Seifert}(2015)}]{bara15}%
  \BibitemOpen
  \bibfield  {author} {\bibinfo {author} {\bibfnamefont {A.~C.}\ \bibnamefont {Barato}}\ and\ \bibinfo {author} {\bibfnamefont {U.}~\bibnamefont {Seifert}},\ }\bibfield  {title} {\bibinfo {title} {Thermodynamic {{Uncertainty Relation}} for {{Biomolecular Processes}}},\ }\href {https://doi.org/10.1103/PhysRevLett.114.158101} {\bibfield  {journal} {\bibinfo  {journal} {Physical Review Letters}\ }\textbf {\bibinfo {volume} {114}},\ \bibinfo {pages} {158101} (\bibinfo {year} {2015})}\BibitemShut {NoStop}%
\bibitem [{\citenamefont {Gingrich}\ \emph {et~al.}(2016)\citenamefont {Gingrich}, \citenamefont {Horowitz}, \citenamefont {Perunov},\ and\ \citenamefont {England}}]{ging16}%
  \BibitemOpen
  \bibfield  {author} {\bibinfo {author} {\bibfnamefont {T.~R.}\ \bibnamefont {Gingrich}}, \bibinfo {author} {\bibfnamefont {J.~M.}\ \bibnamefont {Horowitz}}, \bibinfo {author} {\bibfnamefont {N.}~\bibnamefont {Perunov}},\ and\ \bibinfo {author} {\bibfnamefont {J.~L.}\ \bibnamefont {England}},\ }\bibfield  {title} {\bibinfo {title} {Dissipation {{Bounds All Steady-State Current Fluctuations}}},\ }\href {https://doi.org/10.1103/PhysRevLett.116.120601} {\bibfield  {journal} {\bibinfo  {journal} {Physical Review Letters}\ }\textbf {\bibinfo {volume} {116}},\ \bibinfo {pages} {120601} (\bibinfo {year} {2016})}\BibitemShut {NoStop}%
\bibitem [{\citenamefont {Horowitz}\ and\ \citenamefont {Gingrich}(2020)}]{horo20}%
  \BibitemOpen
  \bibfield  {author} {\bibinfo {author} {\bibfnamefont {J.~M.}\ \bibnamefont {Horowitz}}\ and\ \bibinfo {author} {\bibfnamefont {T.~R.}\ \bibnamefont {Gingrich}},\ }\bibfield  {title} {\bibinfo {title} {Thermodynamic uncertainty relations constrain non-equilibrium fluctuations},\ }\href {https://doi.org/10.1038/s41567-019-0702-6} {\bibfield  {journal} {\bibinfo  {journal} {Nature Physics}\ }\textbf {\bibinfo {volume} {16}},\ \bibinfo {pages} {15} (\bibinfo {year} {2020})}\BibitemShut {NoStop}%
\bibitem [{\citenamefont {Shiraishi}(2021)}]{shir21}%
  \BibitemOpen
  \bibfield  {author} {\bibinfo {author} {\bibfnamefont {N.}~\bibnamefont {Shiraishi}},\ }\bibfield  {title} {\bibinfo {title} {Optimal thermodynamic uncertainty relation in {{Markov}} jump processes},\ }\href@noop {} {\bibfield  {journal} {\bibinfo  {journal} {arXiv:2106.11634 [cond-mat]}\ } (\bibinfo {year} {2021})},\ \Eprint {https://arxiv.org/abs/2106.11634} {arXiv:2106.11634 [cond-mat]} \BibitemShut {NoStop}%
\bibitem [{\citenamefont {{Muratore-Ginanneschi}}\ \emph {et~al.}(2013)\citenamefont {{Muratore-Ginanneschi}}, \citenamefont {{Mej{\'i}a-Monasterio}},\ and\ \citenamefont {Peliti}}]{mura13}%
  \BibitemOpen
  \bibfield  {author} {\bibinfo {author} {\bibfnamefont {P.}~\bibnamefont {{Muratore-Ginanneschi}}}, \bibinfo {author} {\bibfnamefont {C.}~\bibnamefont {{Mej{\'i}a-Monasterio}}},\ and\ \bibinfo {author} {\bibfnamefont {L.}~\bibnamefont {Peliti}},\ }\bibfield  {title} {\bibinfo {title} {Heat {{Release}} by {{Controlled Continuous-Time Markov Jump Processes}}},\ }\href {https://doi.org/10.1007/s10955-012-0676-6} {\bibfield  {journal} {\bibinfo  {journal} {Journal of Statistical Physics}\ }\textbf {\bibinfo {volume} {150}},\ \bibinfo {pages} {181} (\bibinfo {year} {2013})}\BibitemShut {NoStop}%
\bibitem [{\citenamefont {Dechant}(2022)}]{dech22c}%
  \BibitemOpen
  \bibfield  {author} {\bibinfo {author} {\bibfnamefont {A.}~\bibnamefont {Dechant}},\ }\bibfield  {title} {\bibinfo {title} {Minimum entropy production, detailed balance and {{Wasserstein}} distance for continuous-time {{Markov}} processes},\ }\href {https://doi.org/10.1088/1751-8121/ac4ac0} {\bibfield  {journal} {\bibinfo  {journal} {Journal of Physics A: Mathematical and Theoretical}\ }\textbf {\bibinfo {volume} {55}},\ \bibinfo {pages} {094001} (\bibinfo {year} {2022})}\BibitemShut {NoStop}%
\bibitem [{\citenamefont {Ilker}\ \emph {et~al.}(2022)\citenamefont {Ilker}, \citenamefont {G{\"u}ng{\"o}r}, \citenamefont {{Kuznets-Speck}}, \citenamefont {Chiel}, \citenamefont {Deffner},\ and\ \citenamefont {Hinczewski}}]{ilke22}%
  \BibitemOpen
  \bibfield  {author} {\bibinfo {author} {\bibfnamefont {E.}~\bibnamefont {Ilker}}, \bibinfo {author} {\bibfnamefont {{\"O}.}~\bibnamefont {G{\"u}ng{\"o}r}}, \bibinfo {author} {\bibfnamefont {B.}~\bibnamefont {{Kuznets-Speck}}}, \bibinfo {author} {\bibfnamefont {J.}~\bibnamefont {Chiel}}, \bibinfo {author} {\bibfnamefont {S.}~\bibnamefont {Deffner}},\ and\ \bibinfo {author} {\bibfnamefont {M.}~\bibnamefont {Hinczewski}},\ }\bibfield  {title} {\bibinfo {title} {Shortcuts in {{Stochastic Systems}} and {{Control}} of {{Biophysical Processes}}},\ }\href {https://doi.org/10.1103/PhysRevX.12.021048} {\bibfield  {journal} {\bibinfo  {journal} {Physical Review X}\ }\textbf {\bibinfo {volume} {12}},\ \bibinfo {pages} {021048} (\bibinfo {year} {2022})}\BibitemShut {NoStop}%
\bibitem [{\citenamefont {Remlein}\ and\ \citenamefont {Seifert}(2021)}]{reml21}%
  \BibitemOpen
  \bibfield  {author} {\bibinfo {author} {\bibfnamefont {B.}~\bibnamefont {Remlein}}\ and\ \bibinfo {author} {\bibfnamefont {U.}~\bibnamefont {Seifert}},\ }\bibfield  {title} {\bibinfo {title} {Optimality of nonconservative driving for finite-time processes with discrete states},\ }\href {https://doi.org/10.1103/PhysRevE.103.L050105} {\bibfield  {journal} {\bibinfo  {journal} {Physical Review E}\ }\textbf {\bibinfo {volume} {103}},\ \bibinfo {pages} {L050105} (\bibinfo {year} {2021})}\BibitemShut {NoStop}%
\bibitem [{\citenamefont {Yoshimura}\ \emph {et~al.}(2023)\citenamefont {Yoshimura}, \citenamefont {Kolchinsky}, \citenamefont {Dechant},\ and\ \citenamefont {Ito}}]{yosh23}%
  \BibitemOpen
  \bibfield  {author} {\bibinfo {author} {\bibfnamefont {K.}~\bibnamefont {Yoshimura}}, \bibinfo {author} {\bibfnamefont {A.}~\bibnamefont {Kolchinsky}}, \bibinfo {author} {\bibfnamefont {A.}~\bibnamefont {Dechant}},\ and\ \bibinfo {author} {\bibfnamefont {S.}~\bibnamefont {Ito}},\ }\bibfield  {title} {\bibinfo {title} {Housekeeping and excess entropy production for general nonlinear dynamics},\ }\href {https://doi.org/10.1103/PhysRevResearch.5.013017} {\bibfield  {journal} {\bibinfo  {journal} {Physical Review Research}\ }\textbf {\bibinfo {volume} {5}},\ \bibinfo {pages} {013017} (\bibinfo {year} {2023})}\BibitemShut {NoStop}%
\bibitem [{\citenamefont {Van~Vu}\ and\ \citenamefont {Saito}(2023)}]{vu23a}%
  \BibitemOpen
  \bibfield  {author} {\bibinfo {author} {\bibfnamefont {T.}~\bibnamefont {Van~Vu}}\ and\ \bibinfo {author} {\bibfnamefont {K.}~\bibnamefont {Saito}},\ }\bibfield  {title} {\bibinfo {title} {Thermodynamic {{Unification}} of {{Optimal Transport}}: {{Thermodynamic Uncertainty Relation}}, {{Minimum Dissipation}}, and {{Thermodynamic Speed Limits}}},\ }\href {https://doi.org/10.1103/PhysRevX.13.011013} {\bibfield  {journal} {\bibinfo  {journal} {Physical Review X}\ }\textbf {\bibinfo {volume} {13}},\ \bibinfo {pages} {011013} (\bibinfo {year} {2023})}\BibitemShut {NoStop}%
\bibitem [{\citenamefont {Nagayama}\ \emph {et~al.}(2025)\citenamefont {Nagayama}, \citenamefont {Yoshimura},\ and\ \citenamefont {Ito}}]{naga25}%
  \BibitemOpen
  \bibfield  {author} {\bibinfo {author} {\bibfnamefont {R.}~\bibnamefont {Nagayama}}, \bibinfo {author} {\bibfnamefont {K.}~\bibnamefont {Yoshimura}},\ and\ \bibinfo {author} {\bibfnamefont {S.}~\bibnamefont {Ito}},\ }\bibfield  {title} {\bibinfo {title} {Infinite variety of thermodynamic speed limits with general activities},\ }\href {https://doi.org/10.1103/PhysRevResearch.7.013307} {\bibfield  {journal} {\bibinfo  {journal} {Physical Review Research}\ }\textbf {\bibinfo {volume} {7}},\ \bibinfo {pages} {013307} (\bibinfo {year} {2025})}\BibitemShut {NoStop}%
\bibitem [{\citenamefont {van~der Meer}\ and\ \citenamefont {Dechant}(2026)}]{shortpaper}%
  \BibitemOpen
  \bibfield  {author} {\bibinfo {author} {\bibfnamefont {J.}~\bibnamefont {van~der Meer}}\ and\ \bibinfo {author} {\bibfnamefont {A.}~\bibnamefont {Dechant}},\ }\href@noop {} {\bibinfo {title} {Near-optimality of conservative driving in discrete systems}} (\bibinfo {year} {2026}),\ \Eprint {https://arxiv.org/abs/arXiv:2602.18321} {arXiv:2602.18321} \BibitemShut {NoStop}%
\bibitem [{\citenamefont {Risken}(1986)}]{Risken1986}%
  \BibitemOpen
  \bibfield  {author} {\bibinfo {author} {\bibfnamefont {H.}~\bibnamefont {Risken}},\ }\href@noop {} {\emph {\bibinfo {title} {The Fokker-Planck Equation}}}\ (\bibinfo  {publisher} {Springer Berlin},\ \bibinfo {year} {1986})\BibitemShut {NoStop}%
\bibitem [{\citenamefont {Maes}\ and\ \citenamefont {Neto{\v{c}}n{\`y}}(2014)}]{Maes2014}%
  \BibitemOpen
  \bibfield  {author} {\bibinfo {author} {\bibfnamefont {C.}~\bibnamefont {Maes}}\ and\ \bibinfo {author} {\bibfnamefont {K.}~\bibnamefont {Neto{\v{c}}n{\`y}}},\ }\bibfield  {title} {\bibinfo {title} {{A nonequilibrium extension of the Clausius heat theorem}},\ }\href@noop {} {\bibfield  {journal} {\bibinfo  {journal} {Journal of Statistical Physics}\ }\textbf {\bibinfo {volume} {154}},\ \bibinfo {pages} {188} (\bibinfo {year} {2014})}\BibitemShut {NoStop}%
\bibitem [{\citenamefont {Dechant}\ \emph {et~al.}(2022{\natexlab{a}})\citenamefont {Dechant}, \citenamefont {Sasa},\ and\ \citenamefont {Ito}}]{Dechant2022}%
  \BibitemOpen
  \bibfield  {author} {\bibinfo {author} {\bibfnamefont {A.}~\bibnamefont {Dechant}}, \bibinfo {author} {\bibfnamefont {S.-i.}\ \bibnamefont {Sasa}},\ and\ \bibinfo {author} {\bibfnamefont {S.}~\bibnamefont {Ito}},\ }\bibfield  {title} {\bibinfo {title} {Geometric decomposition of entropy production in out-of-equilibrium systems},\ }\href@noop {} {\bibfield  {journal} {\bibinfo  {journal} {Physical Review Research}\ }\textbf {\bibinfo {volume} {4}},\ \bibinfo {pages} {L012034} (\bibinfo {year} {2022}{\natexlab{a}})}\BibitemShut {NoStop}%
\bibitem [{\citenamefont {Dechant}\ \emph {et~al.}(2022{\natexlab{b}})\citenamefont {Dechant}, \citenamefont {Sasa},\ and\ \citenamefont {Ito}}]{Dechant2022a}%
  \BibitemOpen
  \bibfield  {author} {\bibinfo {author} {\bibfnamefont {A.}~\bibnamefont {Dechant}}, \bibinfo {author} {\bibfnamefont {S.-i.}\ \bibnamefont {Sasa}},\ and\ \bibinfo {author} {\bibfnamefont {S.}~\bibnamefont {Ito}},\ }\bibfield  {title} {\bibinfo {title} {Geometric decomposition of entropy production into excess, housekeeping, and coupling parts},\ }\href {https://doi.org/10.1103/PhysRevE.106.024125} {\bibfield  {journal} {\bibinfo  {journal} {Physical Review E}\ }\textbf {\bibinfo {volume} {106}},\ \bibinfo {pages} {024125} (\bibinfo {year} {2022}{\natexlab{b}})}\BibitemShut {NoStop}%
\bibitem [{\citenamefont {Van~Vu}\ and\ \citenamefont {Hasegawa}(2021)}]{vu21}%
  \BibitemOpen
  \bibfield  {author} {\bibinfo {author} {\bibfnamefont {T.}~\bibnamefont {Van~Vu}}\ and\ \bibinfo {author} {\bibfnamefont {Y.}~\bibnamefont {Hasegawa}},\ }\bibfield  {title} {\bibinfo {title} {Geometrical bounds of the irreversibility in {Markovian} systems},\ }\href@noop {} {\bibfield  {journal} {\bibinfo  {journal} {Phys. Rev. Lett.}\ }\textbf {\bibinfo {volume} {126}},\ \bibinfo {pages} {010601} (\bibinfo {year} {2021})}\BibitemShut {NoStop}%
\bibitem [{\citenamefont {Schnakenberg}(1976)}]{schn76a}%
  \BibitemOpen
  \bibfield  {author} {\bibinfo {author} {\bibfnamefont {J.}~\bibnamefont {Schnakenberg}},\ }\bibfield  {title} {\bibinfo {title} {Network theory of microscopic and macroscopic behavior of master equation systems},\ }\href {https://doi.org/10.1103/RevModPhys.48.571} {\bibfield  {journal} {\bibinfo  {journal} {Reviews of Modern Physics}\ }\textbf {\bibinfo {volume} {48}},\ \bibinfo {pages} {571} (\bibinfo {year} {1976})}\BibitemShut {NoStop}%
\bibitem [{\citenamefont {Hill}(1989)}]{hill89}%
  \BibitemOpen
  \bibfield  {author} {\bibinfo {author} {\bibfnamefont {T.~L.}\ \bibnamefont {Hill}},\ }\href {https://doi.org/10.1007/978-1-4612-3558-3} {\emph {\bibinfo {title} {Free Energy Transduction and Biochemical Cycle Kinetics}}}\ (\bibinfo  {publisher} {Springer New York},\ \bibinfo {year} {1989})\BibitemShut {NoStop}%
\bibitem [{\citenamefont {Jiang}\ \emph {et~al.}(2004)\citenamefont {Jiang}, \citenamefont {Qian},\ and\ \citenamefont {Qian}}]{jian04}%
  \BibitemOpen
  \bibfield  {author} {\bibinfo {author} {\bibfnamefont {D.}~\bibnamefont {Jiang}}, \bibinfo {author} {\bibfnamefont {M.}~\bibnamefont {Qian}},\ and\ \bibinfo {author} {\bibfnamefont {M.-P.}\ \bibnamefont {Qian}},\ }\href@noop {} {\emph {\bibinfo {title} {Mathematical theory of nonequilibrium steady-states}}}\ (\bibinfo  {publisher} {Springer Verlag},\ \bibinfo {year} {2004})\BibitemShut {NoStop}%
\bibitem [{\citenamefont {Hatano}\ and\ \citenamefont {Sasa}(2001)}]{hata01}%
  \BibitemOpen
  \bibfield  {author} {\bibinfo {author} {\bibfnamefont {T.}~\bibnamefont {Hatano}}\ and\ \bibinfo {author} {\bibfnamefont {S.-i.}\ \bibnamefont {Sasa}},\ }\bibfield  {title} {\bibinfo {title} {Steady-state thermodynamics of {Langevin} systems},\ }\href@noop {} {\bibfield  {journal} {\bibinfo  {journal} {Phys. Rev. Lett.}\ }\textbf {\bibinfo {volume} {86}},\ \bibinfo {pages} {3463} (\bibinfo {year} {2001})}\BibitemShut {NoStop}%
\bibitem [{\citenamefont {Dechant}\ \emph {et~al.}(2022{\natexlab{c}})\citenamefont {Dechant}, \citenamefont {Sasa},\ and\ \citenamefont {Ito}}]{dech22}%
  \BibitemOpen
  \bibfield  {author} {\bibinfo {author} {\bibfnamefont {A.}~\bibnamefont {Dechant}}, \bibinfo {author} {\bibfnamefont {S.-i.}\ \bibnamefont {Sasa}},\ and\ \bibinfo {author} {\bibfnamefont {S.}~\bibnamefont {Ito}},\ }\bibfield  {title} {\bibinfo {title} {Geometric decomposition of entropy production in out-of-equilibrium systems},\ }\href@noop {} {\bibfield  {journal} {\bibinfo  {journal} {Phys. Rev. Research}\ }\textbf {\bibinfo {volume} {4}},\ \bibinfo {pages} {L012034} (\bibinfo {year} {2022}{\natexlab{c}})}\BibitemShut {NoStop}%
\bibitem [{\citenamefont {Boyd}\ and\ \citenamefont {Vandenberghe}(2004)}]{boyd04}%
  \BibitemOpen
  \bibfield  {author} {\bibinfo {author} {\bibfnamefont {S.}~\bibnamefont {Boyd}}\ and\ \bibinfo {author} {\bibfnamefont {L.}~\bibnamefont {Vandenberghe}},\ }\href@noop {} {\emph {\bibinfo {title} {Convex {Optimization}}}}\ (\bibinfo  {publisher} {Cambridge University Press},\ \bibinfo {year} {2004})\ \bibinfo {note} {google-Books-ID: IUZdAAAAQBAJ}\BibitemShut {NoStop}%
\bibitem [{\citenamefont {Maes}(2020)}]{maes2020a}%
  \BibitemOpen
  \bibfield  {author} {\bibinfo {author} {\bibfnamefont {C.}~\bibnamefont {Maes}},\ }\bibfield  {title} {\bibinfo {title} {Frenesy: {Time}-symmetric dynamical activity in nonequilibria},\ }\href {https://doi.org/10.1016/j.physrep.2020.01.002} {\bibfield  {journal} {\bibinfo  {journal} {Physics Reports}\ }\bibinfo {series} {Frenesy: time-symmetric dynamical activity in nonequilibria},\ \textbf {\bibinfo {volume} {850}},\ \bibinfo {pages} {1} (\bibinfo {year} {2020})}\BibitemShut {NoStop}%
\end{thebibliography}
%

\appendix

\begin{widetext}

\section{Uniqueness of conservative forces for jump processes} \label{app-unique}
We consider a Markov jump dynamics described by the master equation \eqref{master}. 
We focus on rates that can be written in the form \eqref{rates}
\begin{align}
W_t(i,j) = \omega_t(i,j) \exp\bigg[-\frac{\beta}{2} \big( U_t(i) - U_t(j)\big) \bigg] \label{potential-rates},
\end{align}
corresponding to conservative forces $A_t(i,j) = U_t(i) - U_t(j)$.
If we imagine keeping the values $\omega_t(i,j)$ and $U_t(i)$ for some time $t$ fixed, the probabilities $p_{t' > t}(i)$ relax to the instantaneous equilibrium probabilities $\pi_t(i) = e^{-\beta U_t(i)}/\sum_j e^{-\beta U_t(j)}$.
We can then write \eqref{master} as
\begin{align}
d_t p_t(i) = \sum_j  \omega_t(i,j) \Bigg( \sqrt{\frac{\pi_t(i)}{\pi_t(j)}} p_t(j) -  \sqrt{\frac{\pi_t(j)}{\pi_t(i)}} p_t(i) \Bigg)  .
\end{align}
Let us suppose that there exists an alternate choice for the potential $\tilde{U}_t(i)$ and thus $\tilde{p}^\text{eq}_t(i)$, which leads to the same time-evolution of $p_t(i)$ with the same symmetric rates $\omega_t(i,j)$.
Then, we have
\begin{align}
\sum_i  a_t(i) \sum_j  \omega_t(i, j) \Bigg( \Bigg( \sqrt{\frac{\pi_t(i)}{\pi_t(j)}} - \sqrt{\frac{\tilde{\pi}_t(i)}{\tilde{\pi}_t(j)}} \Bigg) p_t(j) -  \Bigg( \sqrt{\frac{\pi_t(j)}{\pi_t(i)}} - \sqrt{\frac{\tilde{\pi}_t(j)}{\tilde{\pi}_t(i)}} \Bigg) p_t(i) \Bigg) = \sum_i a_t(i) \big( d_t p(i) - d_t p(i) \big) = 0.
\end{align}
for arbitrary $a_t(i)$.
Using that $\omega_t(i,j) = \omega_t(j,i)$, we can exchange the indices in the second term and obtain
\begin{align}
0 &= \sum_{i,j}  \omega_t(i, j) \Big(a_t(i) - a_t(j) \Big) \Bigg( \sqrt{\frac{\pi_t(i)}{\pi_t(j)}} - \sqrt{\frac{\tilde{\pi}_t(i)}{\tilde{\pi}_t(j)}} \Bigg) p_t(j)  \\
&= \sum_{i,j}  \omega_t(i, j) \Big(a_t(i) - a_t(j) \Big) \Bigg( \sqrt{\frac{\pi_t(i)}{\tilde{\pi}_t(i)}} - \sqrt{\frac{\pi_t(j)}{\tilde{\pi}_t(j)}} \Bigg) \sqrt{\frac{\tilde{\pi}_t(i)}{\pi_t(j)}} p_t(j) \n .
\end{align}
Now we set $a_t(i) = \sqrt{\pi_t(i)/\tilde{\pi}_t(i)}$, which yields
\begin{align}
0 &= \sum_{i,j}  \Bigg( \sqrt{\frac{\pi_t(i)}{\tilde{\pi}_t(i)}} - \sqrt{\frac{\pi_t(j)}{\tilde{\pi}_t(j)}} \Bigg)^2 \underbrace{\omega_t(i, j) \sqrt{\frac{\tilde{\pi}_t(i)}{\pi_t(j)}} p_t(j)}_{\geq 0} .
\end{align}
If $\omega_t(i,j) > 0$ then this implies
\begin{align}
\frac{\pi_t(i)}{\tilde{\pi}_t(i)} = \frac{\pi_t(j)}{\tilde{\pi}_t(j)} \qquad \Rightarrow \qquad \frac{\pi_t(i)}{\pi_t(j)} = \frac{\tilde{\pi}_t(i)}{\tilde{\pi}_t(j)},
\end{align}
so that $U_i(t) - U_j(t) = \tilde{U}_i(t) - \tilde{U}_j(t)$ and thus the potential $U_i(t)$ is unique up to an additive constant.
If the state space is connected (i.~e.~for every pair of states, there exists a path such that $\omega_t(i,j) > 0$ for every edge $(i,j)$ along the path), then the potential is globally unique up to a constant. 
Otherwise, the potential is unique on every connected component, i.~e., we can add different constants for different connected components.
In both cases, the forces generated by the potential for all transitions with $\omega_t(i,j) > 0$ are unique.
Thus, for fixed $\omega_t(i,j)$, there is a one-to-one correspondence between conservative forces and the time-evolution of the probabilities, provided that the rates can be written in the form \eqref{potential-rates}.

\section{Path ratios for jump processes} \label{app-path}
Here, we derive \eqref{dkl-1} and \eqref{dkl-1} in Section \ref{sec-renorm-path}.
In full generality, we consider two processes parameterized by the rates $W_t^Y(i,j) = e^{Y_t(i,j)} W_t(i,j)$ and $W_t^Z(i,j) = e^{Z_t(i,j)} W_t(i,j)$, where $Y_t(i,j) = -Y_t(j,i)$ and $Z_t(i,j) = -Z_t(j,i)$ are skew-symmetric matrices.
We denote the trajectory probability generated by these processes as $\mathbb{P}^Y(\Gamma)$ and $\mathbb{P}^Z(\Gamma)$, respectively.
The transition probability up to order $dt$ is determined by the transition rates,
\begin{align}
    p^Y(i,t+dt \vert j,t) = \bigg( 1 - \sum_{k} W_t^Y(k,j) dt \bigg) \delta(i,j) + W_t^Y(i,j) dt, \label{transition-short-time}
\end{align}
where we use the convention that $W_t(i,i) = 0$ for all $i$, and similar for the rates $W_t^Z(i,j)$.
Since the dynamics is Markovian, the Kullback-Leibler divergence between trajectory probabilities can be decomposed into a sum of single-transition probabilities, and we can write for the Kullback-Leibler divergence between the time-forward trajectory probabilities,
\begin{align}
    D_\text{KL}\big( \mathbb{P}^Z(\Gamma) \Vert \mathbb{P}^Y(\Gamma) \big) = \sum_{n=1}^N \sum_{i_n} &\sum_{i_{n-1}} \ln \Bigg( \frac{p^Z(i_n,t_n \vert i_{n-1},t_{n-1})}{p^Y(i_n,t_n \vert i_{n-1},t_{n-1})} \Bigg) p^Z(i_n,t_n \vert i_{n-1},t_{n-1}) p^Z_{t_{n-1}}(i_{n-1}) \nn
    &+ \sum_{i_0} \ln \Bigg( \frac{p_0^Z(i_0)}{p_0^Y(i_0)} \Bigg) p_0^Z(i_0) ,
\end{align}
where $t_n = n dt$ and $N = \tau/dt$.
If we assume that the initial state of both processes is the same, then the last term vanishes and it is sufficient to evaluate the Kullback-Leibler divergence between the short-time transition probabilities,
\begin{align}
    D^{Z \Vert Y}_{t_{n-1}} \equiv \sum_{i_n} &\sum_{i_{n-1}} \ln \Bigg( \frac{p^Z(i_n,t_n \vert i_{n-1},t_{n-1})}{p^Y(i_n,t_n \vert i_{n-1},t_{n-1})} \Bigg) p^Z(i_n,t_n \vert i_{n-1},t_{n-1}) p^Z_{t_{n-1}}(i_{n-1}) .
\end{align}
Plugging in \eqref{transition-short-time}, we obtain
\begin{align}
    D^{Z \Vert Y}_{t_{n-1}} = \sum_{i_{n-1}} &\ln \Bigg( \frac{1 - \sum_{k} W_t^Z(k,i_{n-1}) dt }{1 - \sum_{k} W_t^Y(k,i_{n-1}) dt } \Bigg) \bigg( 1 - \sum_{k} W_t^Z(k,i_{n-1}) dt \bigg) p^Z_{t_{n-1}}(i_{n-1}) \\
    &+ \sum_{i_n} \sum_{i_{n-1}} \ln \Bigg( \frac{W^Z_t(i_n,i_{n-1})}{W^Y_t(i_n,i_{n-1})} \Bigg) W^Z_t(i_n,i_{n-1}) p^Z_{t_{n-1}}(i_{n-1}) dt \n .
\end{align}
Expanding the first term for small $dt$ and using the definition of the modified rates, we obtain
\begin{align}
    D^{Z \Vert Y}_{t} = \sum_{j,k} \bigg( e^{Y_t(k,j)} - e^{Z_t(k,j)} + \Big(Z_t(k,j) - Y_t(k,j) \Big) e^{Z_t(k,j)} \bigg) W_t(k,j) p^Z_{t}(j) dt .
\end{align}
Setting $Z_t(i,j) = 0$ and using the parameterization \eqref{param}, we can write this as
\begin{align}
    D^{0 \Vert Y}_{t} = \sum_{j,k} \bigg( e^{Y_t(k,j)} - 1 - Y_t(k,j) \bigg) \Omega_t(k,j) e^{\frac{1}{2} F_t(k,j)} dt .
\end{align}
We now use that $\Omega_t(i,j)$ is symmetric, while $Y_t(i,j)$ and $F_t(i,j)$ are skew-symmetric,
\begin{align}
    D^{0 \Vert Y}_{t} &= \frac{1}{2} \sum_{j,k} \Omega_t(k,j) \Bigg( \bigg( e^{Y_t(k,j)} - 1 - Y_t(k,j) \bigg) e^{\frac{1}{2}  F_t(k,j)} + \bigg( e^{-Y_t(k,j)} - 1 + Y_t(k,j) \bigg) e^{-\frac{1}{2} F_t(k,j)} \Bigg) dt \\
    &=  \sum_{j,k} \Omega_t(k,j) \Bigg( \cosh\bigg( \frac{2 Y_t(k,j) + F_t(k,j)}{2} \bigg) - Y_t(k,j) \sinh\bigg( \frac{F_t(k,j)}{2} \bigg) - \cosh\bigg( \frac{F_t(k,j)}{2} \bigg) \Bigg) dt, \n
\end{align}
which is equal to \eqref{dkl-1}.
Next, we consider the case where the process $Y$ is replaced by its time-reverse, which starts from the final state and undergoes the reverse sequence of transitions under the rates with reverse time-dependence.
We then have
\begin{align}
    D_\text{KL}\big( \mathbb{P}^Z(\Gamma) \Vert \mathbb{P}^{Y \dagger}(\Gamma) \big) = \sum_{n=1}^N \sum_{i_n} &\sum_{i_{n-1}} \ln \Bigg( \frac{p^Z(i_n,t_n \vert i_{n-1},t_{n-1})}{p^Y(i_{n-1},t_n \vert i_{n},t_{n-1})} \Bigg) p^Z(i_n,t_n \vert i_{n-1},t_{n-1}) p^Z_{t_{n-1}}(i_{n-1}) \nn
    &+ \sum_{i_0} \ln \big( p_0^Z(i_0) \big) p_0^Z(i_0) - \sum_{i_N} \ln \big( p_\tau^Y(i_N) \big) p_\tau^Z(i_N) \label{dkl-reverse} .
\end{align}
We first evaluate the contribution from the transitions,
\begin{align}
    D^{Z \Vert Y \dagger}_{t_{n-1}} &\equiv \sum_{i_n} \sum_{i_{n-1}} \ln \Bigg( \frac{p^Z(i_n,t_n \vert i_{n-1},t_{n-1})}{p^Y(i_{n-1},t_n \vert i_{n},t_{n-1})} \Bigg) p^Z(i_n,t_n \vert i_{n-1},t_{n-1}) p^Z_{t_{n-1}}(i_{n-1}) \nn
    &= \sum_{i_{n-1}} \ln \Bigg( \frac{1 - \sum_{k} W_{t_{n-1}}^Z(k,i_{n-1}) dt }{1 - \sum_{k} W_{t_{n-1}}^Y(i_{n-1},k) dt } \Bigg) \bigg( 1 - \sum_{k} W_t^Z(k,i_{n-1}) dt \bigg) p^Z_{t_{n-1}}(i_{n-1}) \nn
    &\qquad + \sum_{i_n} \sum_{i_{n-1}} \ln \Bigg( \frac{W_{t_{n-1}}^Z(i_n, i_{n-1})}{W_{t_{n-1}}^Y(i_{n-1}, i_{n})} \Bigg) W_{t_{n-1}}^Z(i_n, i_{n-1}) p^Z_{t_{n-1}}(i_{n-1}) dt.
\end{align}
Again expanding for small $dt$ in the first term and using the definition of the respective rates, we have
\begin{align}
    D^{Z \Vert Y \dagger}_{t} = \sum_{j,k} \Bigg[ e^{Y_t(j,k) - Z_t(k,j) + \ln \Big(\frac{W_t(j,k)}{W_t(k,j)} \Big)}  - 1 + Z_t(k,j) - Y_t(j,k)  + \ln \Bigg( \frac{W_t(k,j)}{W_t(j,k)} \Bigg) \Bigg] e^{Z_t(k,j)} W_t(k,j)  p_t^Z(j) dt. \label{dkl-reverse-2}
\end{align}
Assuming that $p_\tau^Y(i) = p_\tau^Z(i) = p_\tau(i)$ (for $Z_t(i,j) = 0)$, we can further write the boundary term in \eqref{dkl-reverse} as the time-integral of the Shannon entropy change,
\begin{align}
    \sum_{i_0} \ln \big( p_0(i_0) \big) p_0(i_0) - \sum_{i_N} \ln \big( p_\tau(i_N) \big) p_\tau(i_N) &= -\int_0^\tau dt \ \sum_i \ln \big(p_t(i) \big) d_t p_t(i) \\
    &= - \frac{1}{2} \int_0^\tau dt \ \sum_{i,j} \ln \bigg( \frac{p_t(i)}{p_t(j)} \bigg) \big(W_t(i,j) p_t(j) - W_t(j,i) p_t(i)\big) \n .
\end{align}
Adding this to \eqref{dkl-reverse-2}, we can express the result in term of the parameterization \eqref{param} and use \eqref{entropy-param}. Repeating the same steps as above then results in \eqref{dkl-2}.
    
\end{widetext}

\end{document}